\documentclass[preprint,12pt,eqsecnum]{aastex}
\usepackage{natbib,graphicx}
\newcommand{\be}{\begin{equation}}
\newcommand{\ee}{\end{equation}}

\newcommand{\bea}{\begin{eqnarray}}
\newcommand{\eea}{\end{eqnarray}}

\newcommand{\alphav}{\alpha_*}
\newcommand{\tomega}{\widetilde{\omega}_*}

\begin{document}
\title{Normal Modes of Black Hole Accretion Disks}
\author{Manuel Ortega-Rodr\'{\i}guez\altaffilmark{1}}
\affil{Department of Applied Physics and Gravity Probe B, \\
Stanford University, Stanford, CA 94305--4090; \\
and Escuela de F\'{\i}sica \& Centro de Investigaciones Geof\'{\i}sicas, 
\\
Universidad de Costa Rica, 
San Jos\'e, Costa Rica}
\author{Alexander S. Silbergleit\altaffilmark{2}}
\affil{Gravity Probe B, HEPL, Stanford University, Stanford, CA 94305--4085}
\and\author{Robert V. Wagoner\altaffilmark{3}}
\affil{Department of Physics and \\
Kavli Institute for Particle Astrophysics and Cosmology \\
Stanford University, Stanford, CA 94305--4060}

\altaffiltext{1}{mortega@cariari.ucr.ac.cr}
\altaffiltext{2}{gleit@relgyro.stanford.edu}
\altaffiltext{3}{wagoner@stanford.edu}

\begin{abstract}
This paper studies the hydrodynamical problem of normal modes of small adiabatic oscillations of relativistic barotropic
 thin accretion disks around black holes (and compact weakly magnetic neutron stars). Employing WKB techniques, we obtain 
the eigenfrequencies and eigenfunctions of the modes for different values of the mass and angular momentum of the central black hole. 
We discuss the properties of the various types of modes and show for the first time that modes covering the whole extension of the disk (full--disk p--modes) can exist within the studied thin disk model. However, these modes have relatively short wavelengths.

\end{abstract}


\section{Introduction}

For the last decade our group has been studying the 
hydrodynamical problem of the perturbations of equilibrium models of relativistic accretion disks through a normal-mode formalism. This approach, which we have called diskoseismology, was inspired by the 
work of \citet{kf}, who showed that the effects of general relativity can trap modes near the inner edge of an accretion disk around black holes as well as weakly magnetized neutron stars that are sufficiently compact.
The importance of these 
hydrodynamical
oscillation modes lies in the fact that they provide a potentially powerful probe of both strong gravitational fields and the physics of accretion disks, since most of them do not exist in Newtonian gravity. Their frequencies depend upon the angular momentum as well as the mass of the black hole. (Although we shall not explicitly consider neutron stars here, the results obtained will also apply to them to first order in the 
dimensionless angular momentum parameter $a=cJ/GM^2$, since their exterior metric is identical to that of a black hole to that order. This assumes that the innermost stable orbit of the accretion disk lies outside the neutron star and its magnetosphere.)

While other approaches to the subject involve more local (e.g., dispersion relation) considerations [see, e.g., \citet{k01} and \citet{lubow} and references therein], our focus is instead on  global modes, in which all 
particles oscillate with the same frequency. All of the modes we study are adiabatic, with discrete real eigenfrequencies. However, a discussion of the effects of viscosity is presented at the end of the paper. For reviews of relativistic diskoseismology, see \citet{kfm} and \citet{w}.

The fully relativistic study of normal modes has revealed the existence of 
three basic types of oscillation: g--modes \citep{per}(hereafter RD1), c--modes 
\citep{swo}(hereafter RD2), and p--modes.
The p--modes are pressure-driven oscillations in which (unlike 
the g--modes) the pressure restoring forces act with the gravitational 
restoring forces, making the corotating eigenfrequencies larger than the radial 
epicyclic frequency. 
\citet{osw1}(hereafter RD3) studied the properties of fundamental p--modes with low frequency (those axisymmetric modes with no nodes in their vertical distribution, and with the smallest 
possible eigenfrequency). A summary of our previous results for all three types of modes and their potential relevance in determining the mass and angular momentum of black holes was presented by \citet{wso}.

In the present paper we aim at preparing all the tools needed for the a more general hydrodynamical
analysis by means of an extended WKB method. They can then be applied to studying modes of all types; in this paper, the emphasis is on p--modes, since the WKB technique has not yet been fully used to study them. We investigate various qualitative and quantitative properties of the modes which are axisymmetric as well as non-axisymmetric. They are located near the inner edge of the disk (p-- and c--modes), close to but not including the inner edge (g--modes), in the outer region of the disk (p--modes), as well as covering the whole disk (p--modes). 

In Section 2, the basic formalism that governs our approach is summarized, and the classification of modes is given, including full--disk p--modes never considered before. In Section 3 the WKB solutions of the radial eigenvalue problem are obtained for modes of all types and locations. Section 4 deals with certain general properties of the vertical eigenvalue problem; in particular, it is shown that the number of negative eigenvalues (needed for p--modes) is at most finite, in the proper parameter range (the complete proof is found in the Appendix). The WKB solutions of the vertical eigenvalue problem are given in Section 5. Based on them, the non--existence of both inner and outer high--frequency p--modes is demonstrated; on the other hand, explicit asymptotic solutions for the vertical eigenvalues of g--modes and full-disk p--modes are found. Section 6 contains two applications of the developed techniques, namely, explicit asymptotic formulas for the eigenfrequencies of g--modes and full-disk p--modes are derived. In Section 7, a carefull description of the unperturbed disk model, based on the results of \citet{nt} and \citet{pt}, is given. The model can be used in various studies of accretion disks. Section 8 describes how numerical results are obtained from the formalism, before Section 9 closes the paper with a general discussion.

\section{Basic Assumptions and Equations}

\subsection{Hydrodynamic structure of the unperturbed accretion disk}

We take $c=1$, and express all distances in units of $GM/c^2$ and all frequencies in units of $c^3/GM$ (where $M$ is the mass of the central body) unless otherwise indicated.
The unperturbed disk model and equations for perturbations are exactly the same as in our previous papers RD1, RD2, and RD3, so the account here is very brief.

We employ the Kerr metric to study a thin accretion disk, neglecting its 
self-gravity. The stationary ($\partial/\partial t=0$), symmetric about the 
midplane $z=0$, and axially symmetric ($\partial/\partial\varphi$=0) 
equilibrium disk is taken to be described by the standard relativistic thin
disk model \citep{nt,pt}. The velocity components $v^r=v^z=0$, and the disk 
semi-thickness $h(r)\sim c_s/\Omega\ll r$, where $c_s(r,z)$ is the speed of 
sound. The key frequencies, associated with free-particle orbits, are
\bea \label{3freq}
\Omega(r) & = & (r^{3/2}+a)^{-1}\; , \nonumber \\
\Omega_\perp(r) & = & \Omega(r)\left(1-4a/r^{3/2}+3a^2/r^2\right)^{1/2}\; , 
\nonumber \\
\kappa(r) & = & \Omega(r)\left(1-6/r+8a/r^{3/2}-3a^2/r^2\right)^{1/2}\; 
;\label{eq:1}
\eea
the rotational, vertical epicyclic, and radial epicyclic frequencies, respectively \citep{ag}. The angular momentum parameter $a=cJ/GM^2$ is less than unity in absolute value. 

The inner edge of the disk is at approximately the radius of the last 
stable free-particle circular orbit $r=r_i(a)$, where the epicyclic 
frequency $\kappa(r_i)=0$. This radius $r_i(a)$ is a decreasing function of 
$a$, from $r_i(-1)=9$ through $r_i(0)=6$ to $r_i(1)=1$. So all the 
relations we use are for $r>r_i$, where $\kappa(r)>0$. Note, in particular, 
that
\be
\Omega(r)>\Omega_\perp(r)>\kappa(r) \; ,\quad a>0 \; ;\qquad 
\Omega_\perp(r)>\Omega(r)>\kappa(r) \; ,\quad a<0 \; .
\label{eq:2}
\ee
For $a=0$ (a non-rotating black hole), 
$\Omega(r)=\Omega_\perp(r)>\kappa(r)$. We will also employ the outer disk 
radius, $r_{o}$. For stellar mass black holes in low-mass X-ray binaries, 
$r_{o}\lesssim 10^5 M_\sun/M$ (see RD3 for more details).

We here consider barotropic disks [$p=p(\rho)$, vanishing buoyancy frequency]. (A generalization to small non-zero buoyancy may be found in RD2, Section 4.3, and RD1.) 
In this case hydrostatic equilibrium provides the vertical density and pressure profiles
\be
\rho=\rho_0(r)(1-y^2)^{g}\; ,\quad p=p_0(r)(1-y^2)^{g+1}\;, \quad
g\equiv1/(\Gamma-1) >0 \; , \label{eq:3}
\ee
where $\Gamma>1$ is the adiabatic index (for brevity and convenience, we 
will use parameters $g$ and $\Gamma$ alternatively). One has $\Gamma=4/3$ 
within any radiation pressure dominated region of the disk, and 
$\Gamma=5/3$ within any gas pressure dominated region. The disk surfaces are 
at $y=\pm1$, with $y$ related to the vertical coordinate $z$ by
\be 
y={z\over h(r)}\,\sqrt{\Gamma-1\over2\Gamma} = \frac{\beta\Omega_\perp z}{(2g)^{1/2}c_s(r,0)}  \; , \label{yz}
\ee
and the disk semi-thickness $h(r)$ specified by equation (2.19) of 
RD1.
More information on the unperturbed disk is given in Section 7.

\subsection{Hydrodynamic equations for the disk perturbations}

To investigate the eigenmodes of the disk oscillations, we apply the 
general relativistic formalism that \citet{il} developed for perturbations 
of purely rotating perfect fluids.  Neglecting the self-gravity of the disk 
is usually a very good approximation, examined in Section 7. One can then 
express the 
Eulerian perturbations of all physical quantities through a 
single function $\delta V = \delta p/(\rho\beta\omega)$ which satisfies a 
second-order partial differential equation (PDE). Due to the stationary and 
axisymmetric background, the angular and time dependences are factored out 
as $\delta V = V(r,z)\exp[i(m\phi + \sigma t)]$, where $\sigma$ is the 
eigenfrequency, and the PDE becomes:
\begin{equation}
\frac{\partial}{\partial r}\left(\frac{g^{rr}\rho}{\omega^2-\kappa^2}\frac{\partial V}{\partial r}\right) +
\frac{\partial}{\partial z}\left(\frac{\rho}{\omega^2}\frac{\partial V}{\partial z}\right) + 
\frac{\rho\beta^2}{c_s^2}\,V = 0 \; .
\label{PDE}
\end{equation}
Here $\beta = dt/d\tau = U^0$, the mode corotation frequency 
\[\omega=\omega(r,\sigma)=\sigma+m\Omega(r)\; , \]
and $g^{rr}=1/g_{rr}$ is a Kerr metric component in Boyer-Lindquist coordinates. As in RD1 (at the end of section 2.1), the assumptions $h/r\ll|\omega|/\Omega\ll r/h$ and $|m|\ll r/h$ have been made here to simplify the equation, since $h\ll r$. The terms neglected do not change the results appreciably, except near the corotation resonance. 

Note that the first of these inequalities is violated in a small neighborhood of the corotation radius ($r_c$), where $\omega(r_c)=0$. Since $\Omega(r)$ is a monotonically decreasing function, a corotation resonance exists in the disk (spanning radii $r_i<r<r_o$) if the mode frequency $|\sigma|$ is within the range
\be
 |m|\Omega(r_o)<|\sigma|<|m|\Omega(r_i)\; . \label{cr} 
\ee
Since p--modes are trapped where $\omega^2 > \kappa^2$, the resonance lies where they rapidly vanish. For g--modes ($\omega^2 < \kappa^2$), those of interest (i.~e., excluding large values of $|m|$ and the radial mode number $n$) have $r_c < r_i$ [thus violating the upper limit in equation (\ref{cr})], as can be seen from Figure 3 and the Tables in RD1, as well as equation (\ref{gfr}) below. 

Another assumption of strong variation of modes in the radial direction (characteristic radial wavelength $\lambda_r\equiv |V/(\partial V/\partial r)| \ll r$) ensures the approximate WKB separability of variables in this PDE, $V(r,z) = V_r(r)V_y(r,y)$. The `vertical' part, $V_y$, of the functional amplitude $V(r,z)$ varies slowly with $r$, as do all unperturbed properties of the disk [such as the key frequencies (\ref{eq:1})]. Multiplying equation (\ref{PDE}) by $(\beta^2\rho V_r V_y)^{-1}c_s^2(r,0)$ and using equations (\ref{eq:3}) and (\ref{yz}) with $c_s^2 = dp/d\rho$ gives
\be
-\frac{1}{\alpha^2 V_r}\frac{\partial}{\partial r}\left[\frac{1}{(\omega^2-\kappa^2)}\frac{\partial V_r}{\partial r}\right] 
= \frac{1}{2g\omega_*^2\rho V_y}\frac{\partial}{\partial y}\left(\rho\frac{\partial V_y}{\partial y}\right) 
+ \frac{1}{(1-y^2)} \; , \label{sep}
\ee
where
\be
\omega_*\equiv \omega(r,\sigma)/\Omega_\perp(r) \;  \label{omegastar}
\ee
and 
\be 
\alpha(r) \equiv \beta\sqrt{g_{rr}}/c_s(r,0) \; . 
\label{alpha}
\ee
Since the left-hand side of this equation is a rapidly varying function of $r$ and the right-hand side is a rapidly varying function of $y$, they are both equal to a slowly varying separation function $S(r)$. Then with $\rho^{-1} \partial\rho/\partial y = -2gy(1-y^2)^{-1}$, defining the alternative separation function $\Psi(r) = \omega_*^2(1-S)$ gives
the resulting ordinary differential equations for the vertical ($V_y$) and 
radial ($V_r$) eigenfunctions:
\bea
(1-y^2)\,\frac{d^2V_y}{d y^2} - {2gy}\,\frac{d V_y}{d y} + {2g\omega_*^2}\,
\left[1 -\left(1-\frac{\Psi}{\omega_*^2}\right)\left(1-y^2\right)\right]V_y 
= 0 \; ,
\label{eq:4} \\
\frac{d^2 V_r}{dr^2} - \frac{1}{(\omega^2-\kappa^2)} \left[\frac{d}{dr}
(\omega^2-\kappa^2)\right]\frac{dV_r}{dr} +\alpha^2(\omega^2-\kappa^2)\left(1 -
\frac{\Psi}{\omega_*^2}\right)V_r = 0 \; . \label{eq:5}
\eea
Note that, according to equation (\ref{eq:5}), the radial wavelength 
$\lambda_r$ is approximately expressed as
\be
\lambda_r^{-1}=\alpha(r)\sqrt{(1-\Psi/\omega_*^2)(\omega^2-\kappa^2)} \; . 
\label{lamr}
\ee

As in RD2,  we assume that
\be
\alpha(r)=\gamma(r)\frac{r^{\mu+\nu}}{(r-r_i)^{\mu}} \; , \qquad
0 \leq \mu < 1/2 \; , \quad \nu\geq 0 \; , \label{3.7}
\ee
where $\gamma(r)$ is some function bounded from above and away from zero, 
varying slowly with radius, and tending at infinity to a limit 
$\gamma(\infty) \equiv \gamma_\infty>0$.
The singularity at $r=r_i$ is due to the fact that the pressure at the 
inner edge vanishes in a typical disk model [$p\propto (r-r_i)^k$] if the 
torque does, in which case $\mu=(\Gamma-1)k/2\Gamma$. Typically, $k=2$ and 
$\Gamma=5/3$, giving $\mu=2/5$. If even a small torque is applied to the 
disk at the inner edge, which is likely due to magnetic stress \citep{hk}, 
the pressure and the speed of sound become nonzero at $r=r_i$. This 
corresponds to a nonsingular $\alpha(r)$ with $\mu=0$.

It is also true that  $\nu<1$ in most disk models. If it is less than 
one--half, some limitation on the disk size (given $\gamma_\infty$, and 
vice versa) applies, which follows from (\ref{alpha}) and the thin disk 
requirement of the previous section, that is, $h(r)\sim c_s/\Omega\sim 1/\alpha\Omega\ll r$.
Details can be found in RD3.

Together with the appropriate homogeneous boundary conditions (discussed by 
RD1 and RD2, and below in sections 3 and 4), equations (\ref{eq:4}) and (\ref{eq:5}) generate the vertical and radial eigenvalue problems, respectively, which are entangled through $\sigma$ and $\Psi$ entering both of them. The radial boundary  conditions depend on the type of mode and its capture zone, as discussed in the following sections. The coefficient $\alpha$, the vertical eigenfunction $V_y$ and eigenvalue (separation function) $\Psi$ vary slowly with radius (comparable to the radial variation of the properties of the equilibrium model), as do $\omega(r)$ and $\omega_*(r)$. 

Our goal is to determine the vertical eigenvalues $\Psi(r,\sigma)$ and the 
corresponding spectrum of eigenfrequencies $\sigma$ from the two eigenvalue 
problems for equations (\ref{eq:4}) and (\ref{eq:5}). Along with the 
angular mode number $m$, we employ $j$ and $n$ for the vertical and radial 
mode numbers (closely related to the number of nodes in the corresponding 
eigenfunction), respectively, as in our previous works cited above. 
It is worthwhile to note that the eigenvalue $\Psi$ depends on the parameters $r, \sigma, a,$ and $m$ through one combination only, namely $\omega_*^2$, which is clearly seen from 
equation (\ref{eq:4}).

\subsection{Classes of modes}

\subsubsection{Modes in the Lindblad resonance frequency range}

The classification of the oscillation modes of accretion disks was explicitly discussed in RD2 and RD3, so here we just review it briefly, and make one addition that was missing before. Consider first the eigenfrequency range
\be
\sigma_-<\sigma<\sigma_+\;,
\label{s<}
\ee
where
\be
\sigma_+(m, a)=\max_{r_i<r<r_o}\left[-m\Omega(r)+\kappa(r)\right],\qquad
\sigma_-(m, a)=\min_{r_i<r<r_o}\left[-m\Omega(r)-\kappa(r)\right]\; ,
\label{eq:8}
\ee
which is characterized by the existence of the Lindblad resonances, i.e., the two roots of the equation $\omega^2(r)-\kappa^2(r)=0$, denoted $r_\pm=r_\pm(\sigma, m, a)$,  $r_i<r_-<r_+<r_o$.

P--modes are, defined as those for which $\Psi/\omega_*^2-1<0$; with $\sigma$ in this range, the coefficient in front of $V_r$ in equation (\ref{eq:5}) is positive within $r_i<r<r_-$ and $r_+<r<r_o$ for p--modes. So, in the case (\ref{s<}) two types of p--modes are possible: inner p--modes, with capture (oscillation) zone $r_i<r<r_-$, and outer p--modes trapped within $r_+<r<r_o$. The c--modes are those with the difference $1-\Psi/\omega_*^2$ changing its sign; as shown in RD2, their capture zone is $r_i<r<r_c<r_-$, with $r_c$ closely related to their eigenfrequency,  which is the Lense---Thirring frequency at $r=r_c$. Finally, the g--modes are defined as those for which $\Psi/\omega_*^2-1>0$, which implies that their capture zone is between the Lindblad resonances $r_-$ and $r_+$. This completes the mode classification within the frequency range (\ref{s<}).

Note that, as seen from equation (\ref{eq:8}),
\[
\sigma_\pm(0, a)=\pm \kappa_{max}(a)\equiv \max_{r_i<r<r_o}\,\kappa(r)\;,\qquad
\sigma_\pm(-m, a)=-\sigma_\mp(m, a)\;,
\]
so one can limit oneself to $m\geq0$. Although $\sigma_\pm=\pm\kappa_{max}$ for $m=0$, for $m\geq1$ both $\sigma_+$ and $\sigma_-$ are negative, because $\Omega(r)>\kappa(r)$. Moreover, $|\sigma_+|$ is very small, which is clear immediately from its definition (\ref{eq:8}). Indeed, for $m>0$ the quantity $-m\Omega(r)+\kappa(r)$ is negative and goes rapidly to zero at large radii; therefore $\sigma_+(m,a)=-m\Omega(r_o)+\kappa(r_o)$ and $|\sigma_+(m, a)|\ll1$. The dependence of $\kappa_{max}$ on $a$ can be seen in the first row of Table 1 and also in Figure 5a of RD1. The values of $-\sigma_-$ for $m=1,2$ are given in the second and third rows of Table 1. (The values of $|\sigma_+| \sim 10^{-7}$ for $m\geq2$ and they are even smaller for $m=1$.) The corresponding values of the radii for the quantities in 
Table 1 are found in Table 2. 

\subsubsection{Full--disk p--modes}

In the eigenfrequency range complement to (\ref{s<}), namely, 
\be
\sigma>\sigma_+\;\quad{\rm or}\quad \sigma<\sigma_-\; ,
\label{eq:7}
\ee
one has
\[ 
|\omega(r)|=|\sigma+m\Omega(r)|>\kappa(r)\; , 
\]
hence $(\omega^2-\kappa^2)>0$ throughout the disk, and no g--modes are possible; whether c--modes exist in this range is not yet established, although their existnece seems rather improbable. Still, the whole disk can be a capture zone for some p--modes, which we call full--disk p--modes. These modes seem to skip the investigator attention before.

It is natural to discuss also the full--disk modes whose 
eigenfrequencies lie within the `g--mode' range (\ref{s<}); they clearly do 
exist. This is seen immediately from the radial equation (\ref{eq:5}), or, 
even better, from its transformed equivalent given by equation (\ref{rad}) below, by way of 
analogy with quantum mechanics problems.

However, the absolute majority of such full--disk p--modes have a very 
specific radial distribution. Namely, they are strongly concentrated at 
either the inner or the outer region of the disk, and practically absent in the 
remaining part of it, which means that the modes reduce, in fact, to either 
inner or outer modes (studied by RD3 in the low frequency limit). 
Evidently, the full--disk modes with eigenfrequencies in the range 
(\ref{s<}) oscillate in both the inner, $r_i<r<r_-$, and outer, 
$r_+<r<r_o$, parts of the disk, and tunnel through its central region, 
$r_-<r<r_+$. The length of the latter is huge, unless the mode frequency is 
very close to either $\sigma_+-0$ or $\sigma_-+0$. Therefore, if
(for instance) the amplitudes of the growing and decreasing (through 
$r_-<r<r_+$) components of some full--disk p--mode are of the same order at 
$r=r_-$, then the total mode amplitude is enormously larger in the outer 
region compared to the rest of the disk, so that effectively this p--mode 
turns out to be an outer one. In the opposite case, when the growing and 
decreasing component amplitudes are comparable at $r=r_+$, one gets 
effectively an inner p--mode. All other cases are in fact non-generic 
because they require fine tuning of parameters.

\section{WKB Solution of the Radial Eigenvalue Problem}

In this section we construct radial WKB solutions for p-- and g--modes. The detailed derivation of the corresponding solution for c--modes was given in RD2, section 5. In fact, the radial WKB equation for g--modes was also obtained before, yet in RD1; however, there it was derived from an approximate radial equation, instead of using the exact transformation of section 3.1 first introduced in RD3.

The essential assumption for the validity of the WKB solutions is that the 
corresponding capture zone does not shrink, so that we can well separate 
the vicinities  of the boundary points from the core of the domain where 
the standard WKB approximation works. This assumption fails only for 
low-frequency p--modes with $m=0$; low-frequency fundamental p--modes, for which, 
in addition, $j=0$, have already been studied by RD3.

\subsection{Transformed radial equation and boundary conditions}

We will exploit the radial equation (\ref{eq:5}) in a convenient form 
derived in RD3. Namely, we replace $V_r$ with $W$ according to
\be
W=\frac{1}{(\omega^2-\kappa^2)}\frac{dV_r}{dr},\quad
V_r=-\frac{1}{\alpha^2\left(1 - \Psi/\omega_*^2\right)}\frac{dW}{dr}\; ,
\label{WV}
\ee
where the second equality  is nothing else as the radial equation (\ref{eq:5}) itself. We then differentiate it in $r$, and change the radial variable $r$ to
\be
\tau=\int_{r_*}^r\alpha^2(r^{'})
\left|1 - \frac{\Psi(r^{'})}{\omega_*^2(r^{'})}\right|
\,dr^{'}\;, \qquad
r_{*}={\rm const}\; \label{tau}
\ee
(the absolute value here serves conviniently both the p-- and g--mode cases). In this way, we arrive at the following equation which is to be solved instead of the equation (\ref{eq:5}):
  \be
\frac{d^2W}{d\tau^2}+P(\tau)W=0\; ,\qquad
P(\tau)\equiv\frac{\omega^2-\kappa^2}
{\alpha^2\left(1 - \Psi/\omega_*^2\right)}
\label{rad}
\ee
For the eigenfrequency in the range (\ref{s<}), the coefficient $P(\tau)$ 
is positive in the inner and outer p--mode capture zones, $r_i<r<r_-$ and 
$r_+<r<r_o$, whose boundaries $r_{\pm}$ are the turning points of equation 
(\ref{rad}). For g--modes the capture zone is, as said, $r_-<r<r_+$, with turning points forming its both boundaries. If, on the other hand, the eigenfrequency satisfies 
inequalities (\ref{eq:7}), then $P(\tau)$ is positive throughout the whole 
disk $r_i<r<r_o$, being the capture zone of the full--disk p--modes. 
Expressed via the $\tau$ variable, these capture zones are 
$\tau_i<\tau<\tau_-$, $\tau_+<\tau<\tau_o$, $\tau_-<\tau<\tau_+$, and $\tau_i<\tau<\tau_o$, 
respectively; we use the natural notation $\tau_\lambda=\tau(r_\lambda)$ 
for $\lambda=\pm,\,i,\,o$.

The proper boundary conditions at the disk edges were formulated 
in RD3 as
\be
\omega^2(r_i)\cos \theta_i \; W
+  \sin \theta_i \; \frac{dW}{d\tau} \Biggl|_{\tau = \tau_i} = 0\; , 
\label{3.11}
\ee
\be
\left[\omega^2(r_o)-\kappa^2(r_o)\right]\cos \theta_o \; W
+  \sin \theta_o \; \frac{dW}{d\tau} \Biggl|_{\tau = \tau_o} = 
0\;  \label{3.12}
\ee
  with $-\pi/2<\theta_{i,o}\leq\pi/2$ parametrizing our lack of knowledge
about the physical conditions at the disk edges. 

Equations (\ref{3.11}), (\ref{3.12}) form the complete set of boundary conditions for the radial equation 
(\ref{rad}) in the case of  the full--disk modes. However, for the inner and outer 
p--modes only one of these conditions is employed, since the other boundary 
is formed by one of the turning points,
either $\tau_-$ or $\tau_+$. In both cases we require the solution to decay 
outside the turning point boundary of the trapping zone, that is, for
$\tau > \tau_-$ in the inner and for $\tau < \tau_+$ in the outer case. Consequently, neither of the conditions (\ref{3.11}), (\ref{3.12}) is used at all for g--modes, which are required instead to decay both to the left of $\tau_-$ and to the right of $\tau_+$. The decay requirement concludes the formulation of the radial eigenvalue problem 
for modes of all types (including c--modes, see RD2).

\subsection{Solution for the outer p--modes}

For the outer radial problem the convenient variable is given by
equation (\ref{tau}) with $r_*=r_+$,
\be\label{3.10}
\tau=\int_{r_+}^r\alpha^2(r^{'})
\left[1 - \frac{\Psi(r^{'})}{\omega_*^2(r^{'})}\right]
\,dr^{'}\;, \quad
\tau_{+}=0\;,\quad
\tau_o=\int_{r_+}^{r_o}\alpha^2(r^{'})
\left[1 - \frac{\Psi(r^{'})}{\omega_*^2(r^{'})}\right]
\,dr^{'} \;,
\ee
The radial equation (\ref{rad}) on the interval $0<\tau<\tau_o$ with the 
boundary condition (\ref{3.12}) and the requirement for $W(\tau)$ to decay 
when $\tau$ is negative forms the eigenvalue problem whose WKB solution is 
obtained in a standard way. Namely, for all relevant values of $\tau$ 
except the vicinity of $\tau=0$ we use the WKB expression satisfying 
condition (\ref{3.12}):
\[
W \propto P^{-1/4}(\tau) \cos
\left[
\int^{\tau_o}_\tau P^{1/2}(\tau^\prime) d \tau^\prime -
\arctan\left(k_o\cot\theta_o\right)
\right]\; ;
\]
\be
\label{5.1}
k_o(\sigma,m,a,\Gamma, 
r_o)\equiv\left[{\omega^2(r_o)-\kappa^2(r_o)}\right]{P^{-1/2}(\tau_o)}=
\alpha\sqrt{(1 -\Psi/\omega_*^2)(\omega^2-\kappa^2)}\Biggl|_{r=r_o}\; .
\ee
Recall that in general $\omega_*(r)$ depends also on $\sigma,\,m$ and $a$ 
[via $\Omega = \Omega(r,a)$ and $\Omega_\perp = \Omega_\perp(r,a)$]. 
Also $\alpha(r)$ and the vertical eigenvalue 
$\Psi$ generally depend on the adiabatic index $\Gamma$. This explains the 
dependence of $k_o$ on the parameters. Note that, unlike the standard WKB 
situations, here we cannot automatically neglect the derivative of the 
amplitude $P^{-1/4}(\tau)$, because, according to (\ref{3.7}),
(\ref{tau}) and (\ref{rad}), it grows as some positive power of its 
argument when the value of the radius,  and hence of $\tau$, is large. 
However, the
WKB analysis shows that this term can be neglected if $0.25\,P^{-3/2}\,{dP}/d\tau\ll 1$ at $\tau=\tau_o$, which, by the equalities referred to above, reduces to
\be
{\sigma}/{\Omega(r_o)}\gg{h(r_o)}/{r_o}\; .
\label{5.2}
\ee
Since $\Omega(r_o)\sim r_o^{-3/2}$, this condition is valid except perhaps for
extremely small values of $\sigma$, for which it should be specially checked.

Near the turning point $\tau=0$, as usual, we set
\[
W \propto Ai(-p_+^{1/3} \tau),\qquad p_+=dP(0)/d\tau>0\; ,
\]
where $Ai(z)$ is the Airy function of the first kind. The standard matching 
of this and expression (\ref{5.1}) provides the balance of phases in the form:
\[
\int^{\tau_o}_0 P^{1/2}(\tau) d 
\tau=\pi(n+1/4)+\arctan\left(k_o\cot\theta_o\right)
\;,\qquad n=0,1,2,\dots\;.
\]
(for negative values of $n$  the r.h.s. is negative, so they are discarded). Returning 
here to the integration over the radius according to formula (\ref{3.10}) 
and using equation (\ref{rad}), we arrive at the desired equation for the 
eigenfrequencies:
\be
\int^{r_o}_{r_+} \alpha(r)
\sqrt{(1 -\Psi/\omega_*^2)(\omega^2-\kappa^2)}dr=
\pi(n+1/4)+\arctan\left(k_o\cot\theta_o\right)
\;,\qquad n=0,1,2,\dots\;  \label{5.3}
\ee
This equation contains, in fact, two unknowns, the eigenfrequency $\sigma$ 
and the vertical eigenvalue $\Psi$; the second relation involving them is 
delivered by the vertical eigenvalue problem, sometimes in the form of an 
explicit expression for
$1 -\Psi/\omega_*^2$, as in RD3 for the outer low-frequency 
fundamental p--modes.

\subsection{Solution for the inner p--modes}

  In this case of the inner modes it is convenient to define $\tau$ by 
choosing $r_*=r_i$ in equation (\ref{tau}):
\be
\tau=\int_{r_i}^r\alpha^2(r^{'})
\left[1 - \frac{\Psi(r^{'})}{\omega_*^2(r^{'})}\right]
\,dr^{'}\;, \quad
\tau_{i}=0\;,\quad
\tau_-=\int_{r_i}^{r_-}\alpha^2(r^{'})
\left[1 - \frac{\Psi(r^{'})}{\omega_*^2(r^{'})}\right]
\,dr^{'} \; . \label{3.16}
\ee
The radial eigenvalue problem consists of the equation (\ref{rad}) on the 
interval $0<\tau<\tau_-$ with the boundary condition (\ref{3.11}) and the 
requirement for $W(\tau)$ to decay when $\tau>\tau_-$. As compared to that 
for the outer p--modes above, it is solved in a slightly different manner 
resembling the radial WKB solution for c--modes constructed in Section 5.2 of RD2.

Again, near the turning point $\tau_-$,
\[
W \propto Ai[p_-^{1/3} (\tau_- -\tau)],\qquad p_-=dP(\tau_-)/d\tau<0\; ,
\]
and the WKB approximation for the core of the capture zone matching this is
\be
W \propto P^{-1/4}(\tau) \cos
\left[
\int_{\tau}^{\tau_-} P^{1/2}(\tau^\prime) d \tau^\prime - \pi/4
\right]\; .
\label{5.4}
\ee
However, we cannot make expression (\ref{5.4}) satisfy the boundary 
condition (\ref{3.11}) since the amplitude
$P^{-1/4}(\tau)$ diverges at $\tau=\tau_i=0$ by the force of equalities
(\ref{3.7}), (\ref{tau}) and (\ref{rad}).

Therefore we need one more approximation, for the right vicinity of 
$\tau=0$. According to the equalities last referred to, equation 
(\ref{rad}) in this region becomes, to lowest order:
\be
\frac{d^2W}{d\tau^2}+s_i\tau^{{2\mu}/{(1-2\mu)}}W=0\; ,\qquad
s_i=\omega^2(r_i)(1-2\mu)^{{2\mu}/{(1-2\mu)}}
\left[
\left(1-\Psi/\omega_*^2\right)\biggl|_{r=r_i}\alpha_i^2
\right]^{-{1}/{(1-2\mu)}}\; .\label{5.5}
\ee
Its solution satisfying the boundary condition (\ref{3.11}) is expressed in 
terms of Bessel functions as
\[
W\,\propto 
\tau^{1/2}\left[\omega^2(r_i)\cos\theta_i\frac{\left(\lambda/2\right)^{-b}}{\Gamma_E(1-b)}
J_{b}\left(\lambda\tau^{1/2b}\right)-
\sin\theta_i\frac{\left(\lambda/2\right)^{b}}{\Gamma_E(1+b)}
J_{-b}\left(\lambda\tau^{1/2b}\right) \right] \; ;
\]
\be
  b\equiv\frac{1-2\mu}{2(1-\mu)}\; ;  \qquad
\lambda\equiv s_i^{1/2}\, \left( \frac{1-2\mu}{1-\mu} \right) \; 
.\label{bessel}
\ee
Here $\Gamma_E(z)$ is the Euler gamma--function, and $J_{b}(z)$ is the 
Bessel function of the first kind. Note that $0<b\leq1/2$ for 
$0\leq\mu<1/2$.

After matching the asymptotics of solution (\ref{bessel}) at 
$\tau\to+\infty$ with that of solution (\ref{5.4}) at $\tau\to+0$ we arrive 
at the eigenfrequency equation which, upon returning to the radius as the 
integration variable according to equation (\ref{3.16}), looks similar to 
the result (\ref{5.3}):
\be
\int^{r_-}_{r_i} \alpha(r)
\sqrt{(\omega^2-\kappa^2)(1 -\Psi/\omega_*^2)}dr=
\pi(n+1/2) - \Phi_i
\;,\qquad n=0,1,2,\dots\; , \label{5.7}
\ee
with
\be
\Phi_i \equiv\arctan\left(\frac{Z_+}{Z_-}\tan\frac{\pi b}{2}\right) , \quad
Z_\pm \equiv\frac{\Gamma_E(1-b)}{ \Gamma_E(1+b)} 
\left(\frac{\lambda}{2}\right)^{2b}\sin\theta_i\pm\omega^{2}(r_i)\cos\theta_i 
\; . \label{5.9}
\ee
For the special cases in which $\theta_i = \pi/2, 0$,
\be
\Phi_i = \pm \frac{\pi}{4} \left( \frac{1-2\mu}{1-\mu} \right) \; ,
\label{Fispec}
\ee
respectively.
Again, the above should be solved (for $\sigma$ and $\Psi$) together with 
another equation arising from the vertical eigenvalue problem. Also, if 
applied to the inner low-frequency fundamental p--modes studied 
in RD3, equation (\ref{5.7}) gives a result which is close to 
the one obtained there, despite the fact that the WKB approximation is not 
expected to be valid.

\subsection{Solution for the full--disk p--modes}

Finally, we deal with the frequency range defined in formulas (\ref{eq:7}), 
for which both points $r_-,r_+$ are absent inside the disk. Accordingly, 
the radial equation (\ref{rad}) has no turning points in the trapping region,
that is, in the whole disk. The boundary conditions are (\ref{3.11}) 
and  (\ref{3.12}), so now both boundary parameters $\theta_i$ and 
$\theta_o$ are involved in the problem.

Near the inner edge, exactly as in the previous section, we have solution 
(\ref{bessel}). In the remaining part of the disk the WKB solution 
satisfying the outer boundary condition is valid,
\be
W \propto P^{-1/4}(\tau) \cos
\left[
\int_{\tau}^{\tau_o} P^{1/2}(\tau^\prime) d \tau^\prime - 
\arctan(k_o 
\cot\theta_o)
\right]\; ,
\label{5.10}
\ee
[$k_o$ is defined in (\ref{5.1})]. The standard matching of equations
(\ref{bessel}) and (\ref{5.10}) provides, upon returning to the radial 
variable, the following eigenfrequency equation for full--disk p--modes 
[recall that $\sigma>\sigma_+(m,a)$ or $\sigma<\sigma_-(m,a)$, see Section  2.3]:
\be
\int^{r_o}_{r_i} \alpha(r)
\sqrt{(\omega^2-\kappa^2)(1 -\Psi/\omega_*^2)}dr=
\pi(n+1/4) -\Phi_i+\arctan(k_o \cot\theta_o)
\; , \label{5.11}
\ee
with $\Phi_i$ as in equation (\ref{5.9}).

\subsection{Solution for the g--modes}

The capture zone of g--modes is $r_-<r<r_+$, so it is convenient to set $r_*=r_-$
in the variable definition (\ref{tau}) ,
\be\label{g.1}
\tau=\int_{r_-}^r\alpha^2(r^{'})
\left[ \frac{\Psi(r^{'})}{\omega_*^2(r^{'})}-1\right]
\,dr^{'}\;, \quad
\tau_{-}=0\;,\quad
\tau_+=\int_{r_-}^{r_+}\alpha^2(r^{'})
\left[\frac{\Psi(r^{'})}{\omega_*^2(r^{'})}-1\right]
\,dr^{'} \; .
\ee

The radial equation (\ref{rad}) on the interval $0<\tau<\tau_+$ with the  
requirement for $W(\tau)$ to decay outwards of the capture zone forms yet another eigenvalue problem whose WKB solution is constructed similar to the above ones. Namely, for all values of $\tau$ except those in the vicinity of the boundaries we use the standard WKB representation:
\be\label{g.2}
W \propto P^{-1/4}(\tau) \cos
\left[
\int^{\tau}_0 P^{1/2}(\tau^\prime) d \tau^\prime-\Phi_g 
\right]\; ,
\ee
with some $\Phi_g$. Near the turning point $\tau=0$, as usual, we set
\[
W \propto Ai(-p_-^{1/3} \tau),\qquad p_-=dP(0)/d\tau<0\; .
\]
The standard matching of the two above expressions, similar to the one carried out in section 3.2, shows that ($n^{'},\;n^{''}$ below are integers)
\be\label{g.3}
\Phi_g= \pi (n^{'} - 1/4) \; .
\ee
Near the turning point $\tau=\tau_+$ we set, accordingly,
\[
W \propto Ai[p_+^{1/3} (\tau_+ -\tau)],\qquad p_+=dP(\tau_+)/d\tau>0\; ;
\]
matching this with the expression (\ref{g.1}) in the limit $\tau\to\tau_+-0$, we find the balance of phases to be represented by
\[
\int^{\tau_+}_0 P^{1/2}(\tau) d\tau-\Phi_g= \pi (n^{''} - 1/4)\;, 
\]
or, using the above value of $\Phi_g$,
\[
 \int^{\tau_+}_0 P^{1/2}(\tau) d\tau  = \pi (n + 1/2)  \; , 
\]
where we have denoted $n^{'}+n^{''}=n+1$ (this step is pretty similar to the matching of section 3.3). Returning 
here to the integration over the radius according to formula (\ref{3.10}) 
and using equation (\ref{rad}), we arrive at the desired equation for the 
eigenfrequencies:
\be
\int^{r_+}_{r_-} \alpha(r)
\sqrt{(\Psi/\omega_*^2 - 1)(\kappa^2 - \omega^2)}dr= \pi (n+1/2)
\;,\qquad n = 0,1,2,\dots\;  \label{g.4},
\ee
with the range of the radial mode numbers $n$ specified by the fact that the l.h.s. of this equation is positive. This is nothing else as the equation (4.1) from RDI, with the only difference that the arctangent present in the latter is replaced here by $\pi/2$. The difference stems form the fact that in RDI the eigenvalue problem was considered not over the capture zone $r_-<r<r_+$, but on the whole disk $r_i<r<r_o\approx\infty$. However, for all the eigenfrequencies found in RDI the argument of the arctangent is very large, so that its value practically coincides with $\pi/2$.

All radial WKB solutions, including the one for c--modes from RD2, section 5, are now ready to be used with the proper solutions of the vertical eigenvalue problem. The latter, as menitoned, can be found by any method appropriate in this or that case; in particular, the vertical WKB solutions found below in section 5 can be used.
\section{Properties of the Vertical Eigenvalue Problem and p--modes}

\subsection{Boundary conditions for the vertical eigenvalue problem}

We return now to the vertical equation (\ref{eq:4}) and discuss, first of all, boundary conditions for it. In view of the symmetry conditions
\be
\frac{dV_y}{dy}\Biggl|_{y=0}=0 \qquad{\rm or}\qquad V_y\Biggl|_{y=0}=0
\;  \label{6.1}
\ee
for, respectively, even and odd mode distribution through the disk, we can  
consider only half, $0<y<1$, of the whole interval. Near the singular boundary point $y=1$ 
equation (\ref{eq:4}) can be written as 
\be
\frac{d^2V_y}{d y^2} - \left(\frac{g}{1-y}+\dots\right)\,\frac{d V_y}{d y} +
\left(\frac{g\omega_*^2}{1-y}+\dots\right)\,V_y = 0 \; ,
\label{vertaty=1}
\ee
where the dots stand for regular terms [recall $g=1/(\Gamma-1)$ by equation 
(\ref{eq:3})]. Hence, according to the analytical theory of second order ODEs [see e. g.~\citet{olv}], the general solution to equation (\ref{eq:4}) near $y=1$ has the form:
\be
V_y=C_1v_1(y)+C_2(1-y)^{1-g}v_2(y), \qquad g\not=1\; .
\label{near1}
\ee
Here $C_{1,2}$ are arbitrary constants, and $v_{1,2}(y)$ are certain converging power series of $(1-y)$ with $v_{1,2}(1)=1$.

For  the most physically relevant range of values 
$1<\Gamma<2\quad(g>1)$, the second term in the solution 
(\ref{near1}) goes to infinity at the boundary. Therefore in this 
case the requirement that $V_y(y)$ be finite at $y=1$ implies $C_2=0$ and
specifies thus a unique, up to a constant factor, particular solution of the 
vertical equation (\ref{eq:4}). Hence, it proves to be a proper boundary 
condition for the vertical eigenvalue problem.

However, no fundamental principle forbids one from having 
$\Gamma>2\quad(0<g<1)$. In such case the term with $v_2(y)$ is also finite at the boundary $y=1$ (and even turns to zero there). Therefore  the requirement $|V_y(1)|<\infty$ does not select one of the two particular solutions in the representation (\ref{near1}), hence it cannot serve as a boundary condition to the vertical equation 
(\ref{eq:4}). But, since $1-g<1$, the derivative of this term at $y=1$ is infinite, and we again have to get rid of it, because an infinite derivative is physically unacceptable (see also RD1, Section 2.3).

Finally, for $\Gamma=2\;\; (g=1)$, which is the only remaining point of the physical range $\Gamma>1$, the second particular solution has a logarithmic singularity at $y=1$~[\citet{olv}], and thus once again should be discarded.

The above argument shows that for any $\Gamma>1$ the correct boundary 
condition for the vertical equation (\ref{eq:4}) is that both the 
eigenfunction and its derivative be finite at the boundary, or, 
equivalently, that $V_y(y)$ is regular at $y=\pm 1$. Note that, as observed in RD1 and implied by equation (\ref{vertaty=1}) and the boundary condition, the vertical eigenfunction near the boundary can be written in terms of the Bessel function of the first kind as
\be
V_y(y)\propto (1-y)^{-\xi}J_{2\xi}(2|\omega_*|\sqrt{g(1-y)})\qquad
\xi=(g-1)/2=(2-\Gamma)/{2\left(\Gamma-1\right)}\; .
\label{6.4}
\ee

Note that the physical meaning of the above mathematical conditions is that the Lagrangian pressure perturbation 
should vanish at the disk surface, and its ratio to the unperturbed pressure should at least remain finite there. 
(These are standard boundary conditions.) The complete demonstration of this can be found in section 2.3 of RD1.

\subsection{Lower bound for $\Psi/\omega_*^2$: the number
of eigenvalues corresponding \\
to p--modes is at most finite for $1<\Gamma<2$}

A new fundamental fact about the vertical eigenvalue problem is that, given any $\omega_*^2\geq0$ and $1<\Gamma<2\quad(g>1)$, the spectrum of eigenvalues $\Psi$ is bounded from below. Namely, in this parameter range there exists a bounded function $\Lambda (g,\omega_*^2)$ 
such that for any eigenvalue $\Psi$ the following inequality is valid:
\be
  2g\omega_*^2
\,\left(\frac{\Psi}{\omega_*^2}-1\right)>\,-\Lambda (g,\omega_*^2)\;.
\label{6.5}
\ee
The quantity $\Lambda$ is positive, so p--modes, of course, can exist. 
However, since the spectrum of eigenvalues is discrete, only a finite number of eigenvalues 
for which $\Psi/\omega_*^2-1<0$ can be present. Accordingly, for any given 
$\omega_*$ and $1<\Gamma<2$, the number of vertical eigenfunctions corresponding to p--modes is at most 
finite.

The proof of the inequality (\ref{6.5}) and the definition of $\Lambda$ are 
given in the Appendix. Since the exact expression for $\Lambda$ is too 
complicated for quantitative estimates, a relaxed but explicit 
inequality is also derived there, namely,
\be
\frac{\Psi}{\omega_*^2}>
-\,\frac{g\,\left[{\omega_*^2+(g-1)/2}\right]^2}{2(g-1)^2\omega_*^2}+1\;,\quad
g>1\,\,(1<\Gamma<2)
\; .
\label{6.6}
\ee
For large values of $\omega_*^2$ the magnitude of the r.h.s. grows as $\omega_*^2$, giving thus space for more negative eigenvalues [the same is true for the precise bound $\Lambda (g,\omega_*^2)$ as well].

For $\Gamma\geq2$, no analogs of the inequality (\ref{6.5}) are known, 
so far. However, the WKB approximations of the solutions constructed below 
indicate that the number of vertical p--modes is always finite, 
independent of the value of $\Gamma>1$.

\subsection{Behavior of $\Psi(\omega_*^2)/\omega_*^2$}

For a mode of any kind the value of $\omega_*^2$ changes somewhat, and often significantly, through the mode capture zone. It is important to find out how the ratio $\Psi(\omega_*^2)/\omega_*^2$, and thus the vertical structure of the mode, changes through it.

To check on this, we rewrite the vertical equation (\ref{eq:4}) in the self-adjoint form:
\be
\frac{d}{d y}
\left[
\left(1-y^2\right)^{g}\frac{dV_y}{d y}
\right] +
2g\left[
{\omega_*^2}y^2+\Psi\left(1-y^2\right)
\right]
\left(1-y^2\right)^{g-1}V_y = 0 \; .
\label{vertselfadj}
\ee
We then multiply this by $V_y$ and integrate over the interval $(-1,\,1)$. Using integration by parts and the fact that $V_y(\pm1)$ is finite, we obtain
\[
\int_{-1}^{1}\left(1-y^2\right)^{g}\left(\frac{dV_y}{d y}\right)^2dy=
2g
\,\int_{-1}^{1}\,
\left[
\omega_*^2y^2+\Psi\,\left(1-y^2\right)
\right]
\left(1-y^2\right)^{g-1}V_y^2\,dy\; ,
\]
and thus
\be
\int_{-1}^{1}\,
\left[
\omega_*^2y^2+\Psi\,\left(1-y^2\right)
\right]
\left(1-y^2\right)^{g-1}V_y^2\,dy\geq0\;, 
\label{vertineq}
\ee
which inequality is, of course, trivial when $\Psi\geq0$.

We now assume that the value of $\omega_*^2$ in (\ref{vertselfadj}), and thus of $\Psi$ and $V_y$, is slightly changed, and apply a standard perturbation procedure to determine those changes. In the first order of perturbations we arrive then at
\be
\frac{d}{d\omega_*^2}
\left[
\frac{\Psi(\omega_*^2)}{\omega_*^2}
\right]=-
\frac{\int_{-1}^{1}\,\left[\omega_*^2y^2+\Psi\left(1-y^2\right)\right]\left(1-y^2\right)^{g-1}V_y^2\,dy}
{\omega_*^4\int_{-1}^{1}\,\left(1-y^2\right)^{g}V_y^2\,dy}\leq0\; ,
\ee
where the inequality holds due to (\ref{vertineq}). It shows that, for any values of parameters and any mode, the ratio $\Psi(\omega_*^2)/\omega_*^2$ is a decreasing function of $\omega_*^2$. 

Let us now find out how $\omega_*^2$ itself evolves through the disk, which behavior proves to be far from simplistic, depending on all the parameters ($\sigma, m$ and $a$) involved in
\[
\omega_*(r)= {\sigma}/{\Omega_\perp(r)}+m{\Omega(r)}/{\Omega_\perp(r)}\; .
\]
Note, in particular, that the ratio ${\Omega(r)}/{\Omega_\perp(r)}$ grows with $r$ for $a<0$, is unity for $a=0$, and decreases for $0<a<a_\perp\cong 0.953$. For very rapidly rotating central objects with $a_\perp<a<1$ the ratio is not monotonic, having a single minimum not far from the inner disk edge [see formulas (\ref{eq:1}), (\ref{eq:2}) and RD2]. (For $a>a_\perp$, $\Omega_\perp$ achieves a maximum value within the disk.) The ratio in question is close to unity for all values of $a$ if $r$ is sufficiently large.

So, $\omega_*^2$ grows monotonically through the disk for $m=0$ if $a<a_\perp$. For $m>0$, the same is true in the following cases:
1) $\sigma>0$, $a<0$; 2) $\sigma>0$, not too small, and $0<a<a_\perp$; 3) $\sigma<0$, large enough in  magnitude, and $-1<a<a_\perp$. Otherwise, $\omega_*^2(r)$ is not monotonic through the disk, which is especially clear when $\sigma$ is in the `g--mode' range (\ref{s<}), and the corotation resonance (where $\omega_*=\omega=0$) is present. 

\section{WKB Solution of the Vertical Eigenvalue Problem}

We now prepare for the WKB solution of the vertical eigenvalue problem. 
Near the boundary $y=1$ approximation (\ref{6.4}) will be used, with its 
large argument asymptotics given, to lowest order, by
\be
V_y(y)\propto \frac{
\cos
\left[
2|\omega_*|\sqrt{g(1-y)}-(\pi/2)\left(g-1/2\right)
\right]
}
{(1-y)^{g/2-1/4}}
\; .
\label{6.7}
\ee
To construct approximations for the remaining part of the 
interval  $y\in(0,\,1)$, we recall the self-adjoint form (\ref{vertselfadj}) of the vertical 
equation, and carry out the same kind of change of variable that we used in the case 
of the radial equation to reduce it to the WKB form (\ref{rad}):
\be
\zeta(y)\equiv\int_0^y \left(1-{y^{'}}^2\right)^{-g}\,dy^{'}
\; .
\label{6.8}
\ee
For $g\geq1\,\,(1<\Gamma\leq2)$ the range of the new variable is the 
positive semi-axis. If $0<g<1\,\,(\Gamma>2)$, then $0<\zeta<\zeta_*$, with
\be
\zeta_*=\frac{\sqrt\pi}{2}\frac{\Gamma_E\left(1-g\right)}{\Gamma_E\left(3/2-g\right)} 
\; .
\;
\label{6.9}
\ee
[We set $\zeta_*=\infty$ in the first case $(g \geq 1)$, in order to have 
uniform
notations.] Of course, the inverse function $y=y(\zeta)$ is always well 
defined, and the vertical equation in terms of $\zeta$ is
\be
\frac{d^2V_y}{d\zeta^2}+Q(\zeta)V_y=0\;,\qquad
Q(\zeta)\equiv2g\omega_*^2\,
\left[
1 -
\left(1-\frac{\Psi}{\omega_*^2}\right)\left(1-y^2\right)
\right]
\left(1-y^2\right)^{2g-1}\; .
\label{6.10}
\ee

This equation calls for a WKB treatment if $\omega_*^2\gg1$, or $|\Psi|\gg 1$, or both conditions hold simultaneously. However, the first of them fails at the Lindblad resonances $r_-$ and $r_+$ (when exist) since, by equations (\ref{eq:1}) and (\ref{eq:2}),
\be
\omega_*^{2}(r_\pm)=\kappa^{2}(r_\pm)/\Omega_\perp^{2}(r_\pm)<1\; .
\label{omstatLindb}
\ee
This might not affect the validity of the WKB approximation for the relevant modes (inner and outer p--modes, g--modes), for instance, because the second condition might be still valid there, or for some other reason, suich as just the extended parameter validity typical for various WKB approximations. Note that, as far as the p--modes go, the second condition can apparently be valid only for those of them which have negative vertical eigenvalues. 

\subsection{Solution for g--modes and p--modes with $\Psi>0$}

\subsubsection{Derivation of the WKB equation}

In this case $Q$ is positive within the interval $y\in[0,\,1)$, or 
$\zeta\in[0,\,\zeta_*)$, so, taking into account boundary conditions 
(\ref{6.1}), we can write  the WKB approximation for the whole core of the 
disk as
\[
V_y\propto Q^{-1/4}\cos\left[\phi(y)-{\cal I}\pi/2\right]\;,
\]
\be
\phi(y)=\int_0^{\zeta}Q^{1/2}d\zeta^{'}=|\omega_*|\sqrt{2g}
\int_0^{y}
\left[
1 -\left(1-\frac{\Psi}{\omega_*^2}\right)\left(1-{y^{'}}^2\right)
\right]^{1/2}\left(1-{y^{'}}^2\right)^{-1/2}\,dy^{'}\; ,
\ee
where ${\cal I}=0,1$ for the even or odd eigenfunctions, respectively. For 
$y\to1-0$ to lowest order we have
\[
Q\approx 2^{2g}\,g\omega_*^2
(1-y)^{2g-1}\;,\quad
\mbox{and}\qquad
\phi(y)\approx\phi(1)-2|\omega_*|\sqrt{g(1-y)}\; ,
\]
so the corresponding asymptotic formula for the eigenfunction is
\be
V_y\propto \frac{\cos\left[\phi(1)-2|\omega_*|\sqrt{g(1-y)}-{\cal 
I}\pi/2\right]}
{(1-y)^{g/2-1/4}}\; .
\label{6.11}
\ee
Matching this with the asymptotic expression (\ref{6.7}) of the boundary solution (\ref{6.4}), we 
arrive at the desired equation for the positive eigenvalues:
\be
\int_0^{1}
\sqrt{
\frac{1 -\left(1-\Psi/\omega_*^2\right)\left(1-y^2\right)}
{1-y^2}
}\,dy=\frac{\pi\left(j+\delta_\Gamma\right)}{|\omega_*|\sqrt{2g}}\; ,
\label{6.12}
\ee
where $j$ is any suitable integer, and $\delta_\Gamma$
(introduced in RD1) is:
\be
\delta_\Gamma\,=\,(3-\Gamma)/4(\Gamma-1)\,\, ({\rm even\,\,modes}); \qquad
\delta_\Gamma\,=\,(\Gamma+1)/4(\Gamma-1)\,\, ({\rm odd\,\,modes})\;.
\label{delgam}
\ee

Equation (\ref{6.12}) coincides with the equation (4.2) found in RD1 for the zero buoyancy case, but now it is derived in a more accurate way, because an approximate WKB expression of the vertical eigenfunction for the 
core of the disk was exploited there. Instead, our formula (\ref{6.11}) follows from the full equation (\ref{6.10}) obtained by the substitution (\ref{6.8}).

The l.h.s. of equation~(\ref{6.12}) can be expressed in terms of complete 
elliptic integrals, which proves to be very convenient for the analysis; all the needed information on the elliptic integrals can be found in ~\cite{dw}. 

\subsubsection{Explicit asymptotic solution for g--modes}

By definition, $\Psi/\omega_*^2>1$ for g--modes; due to this the WKB equation (\ref{6.12}) can be esily written in terms of the complete elliptic integral of the second kind, $E(k)$, as
\be
\sqrt{\Psi}\,E\,(\sqrt{1-\omega_*^2/\Psi})={\pi\left(j+\delta_\Gamma\right)}/{\sqrt{2g}}\; .
\label{ellipt-g}
\ee
This is exactly the equation (4.12) of RD1, which has been used there for the effective numerical solution. All the results thus obtained showed that, in fact, $\Psi/\omega_*^2\gg1$ (see Tables 1---3 in RD1). This allows for a simple explicit solution for $\Psi$: indeed, the last inequality implies the limit
\[
k=\sqrt{1-\omega_*^2/\Psi}\to1-0\; ,
\]
in which case, according to the expansion {\bf 774.3} from~\cite{dw},
\[
E(k)=1+0.5\,\left[\ln\left(4/k^{'}\right)-0.5\right](k^{'})^2+\ldots,\qquad k^{'}=\sqrt{1-k^2}=|\omega_*|/\sqrt{\Psi}\to+0\; .
\]
The substitution of this expression in equation ~(\ref{ellipt-g}) provides the following asymptotic solution with the first order correction:
\bea
\sqrt{\Psi}=\frac{\pi\left(j+\delta_\Gamma\right)}{\sqrt{2g}\,E(k)}\approx
\frac{\pi\left(j+\delta_\Gamma\right)}{\sqrt{2g}}\left\{1-0.5\,\left[\ln\left(4/k^{'}\right)-0.5\right](k^{'})^2\right\}\nonumber\\
\approx
\frac{\pi\left(j+\delta_\Gamma\right)}{\sqrt{2g}}
\left[
1-0.5\,\left(\ln\frac{4\sqrt{2g}|\omega_*|}{\pi\left(j+\delta_\Gamma\right)}-0.5\right)
\frac{2g\omega_*^2}{\pi^2\left(j+\delta_\Gamma\right)^2}
\right]\; .
\label{Psi-g}
\eea
This result without the correction in square brackets is given in RD1;  with the correction, it reproduces numerical values of $\Psi$ from Tables 1---3 in RD1 very accurately. Note that the above validity condition for the found solution converts, by the force of the solution itself, into the following inequality
\be
\sqrt{\Psi}/|\omega_*|\approx\frac{\pi\left(j+\delta_\Gamma\right)}{\sqrt{2g}|\omega_*|} \gg1
\label{ellipt-cond}
\ee
Note that to lowest order $\Psi$ does not depend on the radius, i. e., on $\omega_*(r)$.
  
\subsubsection{Solvability criterion and the non-existence of p--modes with $\Psi>0$}

In the range $0<\Psi/\omega_*^2<1$ relevant for the p--modes  the WKB equation (\ref{6.12}) can be reduced to
\be
E\,(\sqrt{1-\Psi/\omega_*^2})=\frac{\pi\left(j+\delta_\Gamma\right)}{|\omega_*|\sqrt{2g}}\; ,
\label{ellipt1}
\ee
with the help of the expressions {\bf 781.05} and {\bf 781.15} from~\cite{dw}.

The minimum possible value, $j_{min}$, of the vertical mode number $j$ is 
specified by the obvious requirement $j + \delta_\Gamma>0$. Since, for both 
mode parities, $\delta_\Gamma$ decreases for $\Gamma>1$, $j_{min}(\Gamma)$ 
is monotonically growing, i. e., larger values of $\Gamma$ are more 
restrictive for the vertical mode number. The standard unperturbed disk 
model (see Section  9) employs $\Gamma=4/3$ in the middle region of the disk 
and $\Gamma=5/3$ in both inner and outer regions. Due to this, at least a 
part of the capture zone of any p--mode, be it an inner, outer, or 
full--disk one, lies in a domain with $\Gamma=5/3$; thus for any p--mode of 
either parity, $j_{min}=j_{min}(5/3)=0$ [by expressions 
(\ref{delgam}), $\delta_{5/3}\,=\,1/2,\,1$ for even and odd modes, 
respectively].

A solvability criterion for (\ref{ellipt1}),
\be
\frac{\sqrt{2g}}{\pi}\,|\omega_*(r)|\,< j+\delta_\Gamma<\, 
\frac{\sqrt{2g}}{2}|\omega_*(r)|\; ,
\label{restrj}
\ee
follows from the inequality $1=E(1)< E(k)< E(0)=\pi/2$ valid for $0< k < 
1$, since the l.h.s. of the equation decreases monotonically. Therefore, a p--mode with $\Psi>0$ and $j\geq j_{min}$ can exist if and only if inequalities (\ref{restrj}) hold throughout its 
whole capture zone. This means, in particular, that, for all the relevant radii, the number $j + \delta_\Gamma$ must be within an interval of the $r$--dependent length $(1/2-1/\pi)\,\sqrt{2g}\,|\omega_*(r)|$, which interval also moves along the $r$ axis. It follows that equation (\ref{ellipt1}) has, in fact, no acceptable solutions at all in the whole disk. 

First, no inner and outer vertical p--modes with $\Psi>0$ are possible in the WKB approximation. This has been proved correct analytically [by examining the solvability criterion (\ref{restrj}) at the Lindblad resonances and showing it fails at least at one of them], and then checking the result numerically. Seecond, the situation with the full--disk modes is not much better: one needs a very special combination of parameters, including the mode 
eigenfrequency, for a single such mode to exist. In this case the reason 
is, essentially, that the mode's trapping zone - the whole accretion disk - 
is too large.

To show the nonexistence, we note that no $\Psi > 0$ full--disk mode with eigenfrequency 
$\sigma$ and angular number $m$ is present if, for  at least one radius 
$r<r_o$,
\be
|\omega_*(r)|<(2/\pi)\,|\omega_*(r_o)|\; .
\label{nofull}
\ee
Indeed, this inequality means that the intervals (\ref{restrj}) taken at 
the  outer edge and at some other point of the disk have an empty 
intersection. Clearly, (\ref{nofull}) is true for $m=0$, when it reduces to 
just $\Omega_\perp(r_o)\approx r_o^{-3/2}<(2/\pi)\Omega_\perp(r)$.

If $m\geq1$, inequality (\ref{nofull}) holds for both positive, 
$\sigma>0>\sigma_+$, and large enough negative, $0>\sigma_->\sigma$, 
eigenfrequencies (see Section  2.3). Depending on the range of $a$, one  
takes either $r=r_i$ or $r=r_\perp$ and exploits the fact that 
$\Omega_\perp(r_o)$ is very small. Thus the only case 
when the full--disk $\Psi > 0$ modes seem possible is the one with the 
negative and extremely small eigenfrequency, $0>\sigma\approx-m\Omega(r_o)\approx -mr_o^{-3/2}>\sigma_+$,
so that $\omega(r_o)$ and $\omega_*(r_o)$ are both very close to zero. But then the applicability condition of the WKB approximation is violated. Therefore generically the full--disk p--modes with 
$\Psi>0$ do not exist (within our WKB approximation). Our numerical results confirm this conclusion.

\subsection{Solution for p--modes with $\Psi<0$}

\subsubsection{Derivation of the WKB equation}

In this case the coefficient $Q$ of equation (\ref{6.10}) turns to zero at 
the point
\be
y_+=\left(1-\omega_*^2/\Psi\right)^{-1/2}=\left(1+\omega_*^2/|\Psi|\right)^{-1/2}\in(0,\,1)\; 
,
\label{y+}
\ee
with $Q<0$ for $0\leq y<y_+$, and $Q>0$ for $y_+<y<1$; we write 
$\zeta_+=\zeta(y_+)$, from equation (\ref{6.8}). The quantity 
$y_+=y_+(r,\sigma,m,a)$ is a convenient parameter which we use freely in 
the sequel instead of and along with the eigenvalue 
$\Psi/\omega_*^2=-y_+^2/(1-y_+^2)$.

Because of the turning point $\zeta_+$, we now need to explore four 
different approximations of the solution to the vertical equation 
(\ref{6.10}), namely:

1) For $0\leq\zeta<\zeta_+$ and away from $\zeta_+$, the standard 
non-oscillating WKB approximation is used,
\be
V_y \propto |Q|^{-1/4}\cosh\Xi \quad {\rm or}\quad|Q|^{-1/4} \sinh\Xi,\qquad \Xi(\zeta)=\int_0^{\zeta}
|Q(\zeta^{'})|^{1/2} d\zeta^{'} \; , 
\label{belowy+}
\ee
for even and odd modes, respectively. When $\zeta\to \zeta_+-0$, its asymptotics is
\bea
V_y\propto
|Q|^{-1/4} \left\{
A \exp[-(2/3)Q_+^{1/2}(\zeta_+ - \zeta)^{3/2}] \pm
A^{-1} \exp[(2/3)Q_+^{1/2}(\zeta_+ - \zeta)^{3/2}]
\right\}
\; ;
\label{as1}\\
Q_+= \frac{dQ(\zeta_+)}{d\zeta}>0\; ,\qquad
A=\exp\left[\int_{0}^{\zeta_+}|Q(\zeta)|^{1/2}\,d\zeta\right]>1\; .
\label{Q+A}
\eea

2) In the vicinity of the turning point $\zeta=\zeta_+$ the standard 
approximation containing a linear combination of both Airy functions is 
used 
\[
V_y\propto Ai[-Q_+^{1/3}(\zeta-\zeta_+)] +K Bi[-Q_+^{1/3}(\zeta-\zeta_+)]\; ,
\]
with $K=$const. Its left far-field ($\zeta-\zeta_+\to-\infty$) asymptotics ,
\[
V_y\propto
(\zeta_+ - \zeta)^{-1/4}
\left\{\frac{1}{2}\exp[-(2/3)Q_+^{1/2}(\zeta_+ - \zeta)^{3/2}]
+K\exp[(2/3)Q_+^{1/2}(\zeta_+ - \zeta)^{3/2}]\right\}\; ,
\]
matches to (\ref{as1}) giving
\be
K=\pm\,1/(2A^2)\;
\label{K}
\ee
for the even and odd modes, respectively. The right far-field 
($\zeta-\zeta_+\to+\infty$) asymptotic formula then reads:
\be
V_y\propto
(\zeta - \zeta_+)^{-1/4}
\cos\left[
(2/3)Q_+^{1/2}(\zeta - \zeta_+)^{3/2} - 3 \pi / 4 \mp \arctan (2A^2)
\right]\; .
\label{as2}
\ee

3) For $\zeta_+<\zeta<\zeta_*$ and away from both ends of this interval, the 
standard oscillating WKB approximation with some phase $\Phi$ is used,
$
V_y\propto
Q^{-1/4}
\cos\left[
\int_{\zeta_+}^{\zeta} Q^{1/2} d\zeta^{'} + \Phi
\right]
$, whose asymptotics at $\zeta\to \zeta_++0$ is described by
\[
V_y\propto
(\zeta - \zeta_+)^{-1/4}
\cos\left[
(2/3)Q_+^{1/2}(\zeta - \zeta_+)^{3/2} + \Phi
\right]\; .
\]
Matching this with (\ref{as2}) provides the value of the phase:
\be
\Phi= - 3 \pi /4 \mp \arctan(2A^2)\;.
\label{Phi}
\ee
The behavior at the opposite end, $\zeta\to \zeta_*-0$, is found to be
\be
V_y\propto
Q^{-1/4}
\cos\left[
\int_{y_+}^yQ^{1/2} (1-{y^{'}}^2)^{-g} dy^{'} + \Phi
\right]\; .
\label{as3}
\ee

4) Finally, near the right boundary $\zeta=\zeta_*$ ($y=1$), solution 
(\ref{6.4}) is used, as in Section  6.1. Its left far-field asymptotics 
(\ref{6.7}), when matched with (\ref{as3}) taking (\ref{Phi}) into account, 
leads to the desired eigenvalue equation:
\bea
\int_{y_+}^1
\sqrt{
\frac{1 -\left(1-\Psi/\omega_*^2\right)\left(1-y^2\right)}
{1-y^2}
}\,dy
=\frac{\pi\left(j+\Delta_\Gamma\right)}{|\omega_*|\sqrt{2g}}\; ,
\label{Psineg}\\
 \Delta_\Gamma\equiv(g+1)/2\pm (1/\pi)\arctan(2A^2)\; .
\label{DeltaG}
\eea
Here the positive sign corresponds to even, and the negative sign to odd
modes. The parameter $A$ defined in equation (\ref{Q+A}) can be expressed 
in terms of complete elliptic integrals of the first and second kind:
\be
A(y_+)=\exp\left\{|\omega_*|\,\sqrt{2g}\,(1-y_+^2)^{-1/2}\left[E(y_+)-(1-y_+^2)K(y_+)\right]\right\}\; 
\label{Aell}
\ee
[note that $A\to1-0$ when $y_+\to+0$ ($\Psi\to-0$), and $A\to+\infty$ when 
$y_+\to1-0$ ($\Psi\to-\infty$)].

\subsubsection{Solvability criterion and the non-existence of inner and outer p--modes with $\Psi>0$}

As before, the l.h.s.~of the eigenvalue equation (\ref{Psineg}) can be 
conveniently written via elliptic integrals:
\be
\frac{E(\sqrt{1-y_+^2})-y_+^2 K(\sqrt{1-y_+^2})}{\sqrt{1-y_+^2}}
=\frac{\pi\left(j+\Delta_\Gamma\right)}
{|\omega_*|\,\sqrt{2g}}\; ,
\; .
\label{ellipt2}
\ee
where $K(k)$ is the complete elliptic integral of the first kind.There is still a very important difference between this equation and equation (\ref{ellipt1}). Unlike the latter, here the l.h.s.~can be infinitesimally small. Indeed, it decreases monotonically in the interval $0<y_+<1$ to zero at its right end, in partucular,
\be
k^{-1}\left[E(k)-(1-k^2)K(k)\right]=(\pi k/4)\,\left[1+k^2/8+{\cal 
O}(k^4)\right]\; ,
\qquad k\to0\; ;
\label{E-K}
\ee
the maximum value of the l.h.s.~at $y_+=0$ is unity.

Therefore, given $\sigma$ and $m$, the $j$th vertical p--mode with $\Psi<0$ 
exists if and only if, for all the pertinent radii,
\be\label{condition-omega}
0<\,j+\Delta_\Gamma\,<
\,{\sqrt{2g}}\,|\omega_*(r)|/{\pi}\; .
\label{restrj2}
\ee
The value of $j_{min}$ is apparently determined by the left inequality here.
In the same way as for the modes with $\Psi>0$, the strongest restriction comes from 
$\Gamma=5/3$ and yields $j_{min}=-1$ for even and $j_{min}=0$ for odd modes.
Otherwise, requirement (\ref{restrj2}) is essentially less restrictive than 
(\ref{restrj}) for the modes with $\Psi>0$. The reason is that all the 
intervals (\ref{restrj2}), for any values of the radius, have 
a common left end at zero. Hence, as soon as $\omega_*(r)\not=0$ throughout 
the mode trapping zone [i.e., $\sigma\not=-m\Omega(r)$ there], the 
intersection of all the intervals (\ref{restrj2}) is not empty: it is the 
shortest interval corresponding to some radius $r=r_{min}$ such that 
$|\omega_*(r_{min})|$ is minimal. If this length satisfies
\be
{\sqrt{2g}}\,|\omega_*(r_{min})|/{\pi}\,>\,j+\Delta_\Gamma\; ,
\label{suff}
\ee
then at least one vertical p--mode with $\Psi<0$ (and $j=j_{min}$) exists.

Nevertheless, it turns out there are no solutions corresponding to either inner or outer p--modes [governed by the WKB equation (\ref{ellipt2})]. This is once again proved analytically same way as before, and then checked numerically.

\subsubsection{Explicit asymptotic solution for full--disk p--modes with $\Psi<0$}

On the other hand, the existence of full--disk (WKB) p--modes with $\Psi<0$ is 
immediately seen. In fact, the existence criterion (\ref{restrj2}) requires 
only that $|\omega_*(r)|$ be large enough throughout the capture zone, 
i.e., the whole accretion disk. But, since the eigenfrequencies in the 
full--disk mode range (\ref{eq:7}) are not bounded, this can be clearly 
achieved by choosing a sufficiently large value of $\sigma$. This argument is supported by the following asymptotic solution of the vertical WKB equation 
(\ref{Psineg}) in the limit $\omega_*^2\gg1$ and $-{\Psi}/{\omega_*^2}\gg 1$.

We have shown in Section 4.3 that the ratio $\Psi/\omega_*^2$
is a decreasing function of $\omega_*^2$.  We will now explore
further properties of this function based on the equation (\ref{Psineg}).
First, we define
\be
\tomega \equiv 
\frac{|\omega_*|\sqrt{2g}}{\pi\left(j+\Delta_\Gamma\right)} \; ,
\ee
and realize that $1-\Psi/\omega_*^2$ depends on $\Gamma$, $j$ and
$\omega_*=\omega_*(\sigma, m, a, r)$ only through the
combination $\tomega$. 
(Even though, according to (\ref{DeltaG}), $\Delta_\Gamma$ depends formally on $1-\Psi/\omega_*^2$ through $\arctan\left[2A^2(y_+)\right]$, 
it turns out that in the range of interest $A\to\infty$, $\arctan\left(2A^2\right)\to\pi/2$, so no such dependence is left.) One can thus use the vertical WKB equation (\ref{Psineg}) to compute what we call a 
`universal curve'  ($\tomega^2$ versus $1 - \Psi/\omega_*^2$), which is shown in Figure 1. The curve does have the predicted monotonic behavior.
 
However, if both $\omega_*^2$ and ($-\Psi/\omega_*^2$) have large  enoguh values,  the vertical WKB equation (\ref{Psineg}) can be solved asymptotically, to give the following two--term expression:
\bea
1 - \Psi/\omega_*^2 = \omega_*^2/K^g_j  + 1/4\equiv(\pi\tomega / 4)^2+1/4, \label{univas}\\
K^g_j \equiv(8/g)\left(j+\Delta_\Gamma\right)^2=(8/g) \left[ j + (g+1 \pm 1)/2\right]^{2} \label{Kjg}\; .
\eea
This two-term asymptotics is also plotted in Figure 1. One would expect the two curves, the exact universal curve and the asymptotic one, to be close for large enough values of $1-\Psi/\omega_*^2$. However, they turn to be in a very good agreement within a remarkably wider range: even for the case 
$- \Psi/\omega_*^2 = 1,\;(\tomega \approx 1.7)$, the two curves differ by less than 2 \%. This means that we can use the analytical formula (\ref{univas}) practically for all the full--disk p--modes treated within the WKB approach, since it begins to fail `much later' than the WKB approximation itself.

\subsubsection{Non--existence of p--modes with $\Psi$ changing sign}

One should mention here yet another possibility to find some p--modes, namely, those whose eigenvalue $\Psi$ changes sign somewhere within the capture zone, so it satisfies equations (\ref{ellipt1}) or (\ref{ellipt2}) on either side of the point $r=r_*$ where it vanishes. (Note that the results of Section 4.3 show that such a change of sign can typically occur just once.)  For inner or outer modes, the simultaneous validity of both solvability criteria at $r=r_*$ restricts the solutions to the odd modes with $j=-1$ only, and then the proper solvability criterion cannot be fulfilled at either $r=r_-$ or $r=r_+$. An analysis of the full--disk modes demonstrates that this is possible only for even modes with $\Gamma=5/3$, $j\gg 1$, and in a very narrow range of radii which may actually be in the $\Gamma=4/3$ region. Thus at most a negligible number of p--modes with a changing sign of $\Psi$ exist within the framework of our WKB approach.

\section{Applications: Explicit Formulas for the Eigenfrequencies\\
 of g--Modes and Full--Disk p--Modes
with $\Psi<0$}

\subsection{Eigenfrequencies of g--modes}

In RDI it was shown that, for given $m\geq0$ and the radial mode number $n$, the g--mode eigenfrequency $\sigma_{mnj}<0$ tends to the frequency range boundary $\sigma_-(m,a)\equiv-\sigma_m$ defined in (\ref{eq:8}) when the vertical mode number $j\to\infty$, i.e., when the eigenvalue $\Psi_j$ becomes large. In fact, all the eigenfequencies from RDI, Tables 1--3, have $\Psi/\omega_*^2\gg1$. In this limit, we can write the radial equation (\ref{g.4}) as
\[
\int^{r_+}_{r_-} \alpha(r)
\sqrt{(\Psi_j/\omega_*^2 - 1)(\kappa^2 - \omega^2)}dr\approx 
\int^{r_+}_{r_-} \alpha(r)\,\sqrt{\Psi/\omega_*^2}\,
\sqrt{\kappa^2 - \omega^2}dr= \pi (n+1/2)
\;,
\]
and then use the explicit asymptotic solution for $\Psi$ found in section 5.1.2. To lowest order, formula (\ref{Psi-g}) from there provides 
\be
\int^{r_+}_{r_-} \frac{\alpha(r)}{\omega_*(r)}
\sqrt{\kappa^2 - \omega^2}dr= \sqrt{2g}\left(\frac{n+1/2}{j+\delta_\Gamma}\right),\qquad n=0,1,\dots ; 
\quad j+\delta_\Gamma>0 
\;,\label{geff}
\ee
with the r.h.s. assumed small ($n$ bounded). The assumption is exactly consistent with the studied limit, in which the capture zone is shrinking, with $r_+$ and $r_-$ both tending to merge at the point $r_m$ where $m\Omega(r)+k(r)$ has its maximum, i.e.,
\[
\frac{d }{dr}\left[m\Omega(r)+\kappa(r)\right]\biggl|_{r=r_m}=0,\qquad 
m\Omega(r_m)+\kappa(r_m)\equiv m\Omega(r_m)+\kappa_m=-\sigma_-(m,a)=\sigma_m\; .
\]
Because of this, to lowest order we can take the various quantities under the integral in the equation (\ref{geff}) to be:
\bea
\alpha(r)\approx\alpha(r_m)\equiv\alpha_m,\quad 
\omega_*(r)\approx\omega_*(r_m)=\kappa (r_m)/\Omega_\perp(r_m)\equiv \kappa_m/\Omega_{\perp m}\; ;\quad\qquad\nonumber \\ 
\kappa(r)-\omega(r)\approx \kappa(r_m)-[\sigma_{mnj}+m\Omega(r_m)]\approx \kappa_m+\sigma_{m}-m\Omega(r_m)]=2\kappa_m\; ;\qquad\label{auxg} \\
\kappa(r)+\omega(r)=\kappa(r)+[\sigma_{mnj}+m\Omega(r)]\approx\sigma_m+\sigma_{mnj}-\lambda(r-r_m)^2\equiv\lambda(r_+-r)(r-r_-) \; ,\nonumber
\eea
where
\be
\lambda=\lambda(m,a)=-\frac{1}{2}\,\frac{d^2 }{dr^2}\left[m\Omega(r)+\kappa(r)\right]\biggl|_{r=r_m}\; ;\label{lamg}
\ee
note that the identity in (\ref{auxg}) implies, in particular, that $r_+-r_-=2\sqrt{(\sigma_m+\sigma_{mnj})/\lambda}$.

Using (\ref{auxg}) and (\ref{lamg}), with the same accuracy as above we find the following value of the integral (\ref{geff}):
\bea
\int^{r_+}_{r_-} \frac{\alpha(r)}{\omega_*(r)}
\sqrt{\kappa^2 - \omega^2}dr\approx
\frac{\alpha_m\Omega_{\perp m}}{\kappa_m}\sqrt{2\kappa_m\lambda}\,\int^{r_+}_{r_-}\sqrt{(r_+-r)(r-r_-)}dr=\nonumber\\
\frac{\alpha_m\Omega_{\perp m}}{\kappa_m}\sqrt{2k_m\lambda}(r_+-r_-)^2\int^{1}_{0}\sqrt{t(1-t)}dt
=\frac{\alpha_m\Omega_{\perp m}\pi}{\sqrt{2\kappa_m\lambda}}\,(\sigma_m+\sigma_{mnj}) 
\;.\label{gint}
\eea
From formulas (\ref{geff}) and (\ref{gint}) we obtain the desired explicit expression for  the g--mode eigenfrequency $\sigma_{mnj}<0$:
\be
\sigma_m+\sigma_{mnj}=\sigma_m-|\sigma_{mnj}|\approx 
\frac{2\sqrt{g\kappa_m\lambda}(n+1/2)}{\alpha_m\Omega_{\perp m}\pi(j+\delta_\Gamma)}=
\sqrt{\frac{2\kappa_m\lambda}{\Psi_j}}\,\frac{n+1/2}{\alpha_m\Omega_{\perp m}}
\;.\label{gfr}
\ee
Up to a notation ${\cal D}_2=2k_m\lambda$, the last formula coincides with the result (4.8) of RDI, where, however, no explicit value of $\Psi$ was given. In addition, the current derivation provides clear asymptotic validity conditions of the result.
  
\subsection{Eigenfrequencies of full--disk p--modes}

\subsubsection{Derivation of the eigenfrequency expressions}

We proceed here in the way similar to the one we followed in the previous case of g--modes. Assuming $\omega_*^2$ and $-\Psi/\omega_*^2$ large enough, we introduce the result (\ref{univas}) to the proper radial WKB equation (\ref{5.11}), which yields a relation:
\be
\int^{r_o}_{r_i} \alpha \Omega_\perp
\sqrt{(\omega_*^2-\kappa^2/\Omega_\perp^2)(\omega_*^2/K^g_j + 1/4 )}dr=
\pi(n+1/4) -\Phi_i+\arctan(k_o \cot\theta_o)
\; .
\label{condition}
\ee
Since $\omega_*^2$ is large, the term $\arctan(k_o \cot\theta_o) \rightarrow \pi/2$ by  formula (\ref{5.1}), unless $\theta_o = \pi/2$, in which case it vanishes; in any case, the term is within  the range $[-\pi/2,\;\pi/2]$. The same is true for the phase shift $\Phi_i$, defined by the formulas (\ref{5.9}), (\ref{bessel}) and (\ref{5.5}), which depends on the inner edge boundary condition parameter $\theta_i$. Hence these terms, as well as the additional $\pi/4$, are very small compared to $\pi n\gg 1$ [from equation (\ref{nlb})]. For this reason, we neglect them in the following results, which become thus independent of the boundary conditions within the whole range of parameters $\theta_i,\;\theta_o$.

Now, upon expanding the square root on the right of equation (\ref{condition}) in inverse powers of $\omega_*^2$, the l.h.s.~of equation (\ref{condition}) becomes
\[
\int^{r_o}_{r_i} \frac{\alpha\Omega_\perp}{(K^g_j)^{1/2}} \;  
\left[ \omega_*^2 +
\frac{1}{2} \left( \frac{K^g_j}{4} - \frac{\kappa^2}{\Omega_\perp^2} \right)
+ O\left(\frac{1}{\omega_*^{2}}\right) \right] dr \; .
\]
Therefore, neglecting the higher order terms, we obtain finally the following analytical
expressions for the eigenfrequencies:
\be
\sigma_{0nj} =  \sqrt{\left(\pi n- I_{3j}\right)/I_{1j}}
\; ;
\label{sigma0}
\ee
\begin{eqnarray}
\sigma_{mnj}^{\pm} & = & m \left\{
\pm \sqrt{\epsilon_j^2+\left(\pi n - I_{3j}-m^2I_{4j}\right)/m^2I_{1j}} -\epsilon_j
\right\}\; ,
\label{sigmam}
\end{eqnarray}
where
\be
\epsilon_j\equiv \frac{I_{2j}}{I_{1j}}\; ,\qquad
I_{1j} \equiv \int_{r_i}^{r_o} \frac{\alpha }{\Omega_\perp(K^g_j)^{1/2}} dr\; ,\qquad 
I_{2j} \equiv\int_{r_i}^{r_o} \frac{\alpha\Omega}{\Omega_\perp(K^g_j)^{1/2}}dr  \; ;
\label{epsI12}
\ee
\be
I_{3j} \equiv \int_{r_i}^{r_o} \frac{\alpha \Omega_\perp}{2(K^g_j)^{1/2}}\,
 \left( \frac{K^g_j}{4} - \frac{\kappa^2}{\Omega_\perp^2} \right)dr\; ,\qquad
 I_{4j} \equiv \int_{r_i}^{r_o} \, 
\frac{\alpha\Omega^2}{\Omega_\perp(K^g_j)^{1/2}} dr \; .
\label{I34}
\ee

Of course, only those values of the mode numbers $m,n,$ and $j$ should be considered which give real values of eigenfrequencies in the proper frequency range (\ref{eq:7}), 
\[
\sigma_{mnj}^{+}>\sigma_{+}(m,a) \; ,\qquad \sigma_{mnj}^{-}<\sigma_{-}(m,a) \; .
\]
It turns out, though, that all the mentioned requirements are not at all restrictive for the modes with the lowest possible radial numbers; the real restrictions stemming from the validity of our approach are discussed in Section 8.

Note that the relative magnitudes of the integrals $I_{ij}$ involved in equations (\ref{sigma0}) and (\ref{sigmam}) are governed mostly by the quantities $\Omega_\perp(r)$ and $\Omega(r)$, which rapidly decrease away from the inner disk edge. 
One thus expects $I_{1j}$ to be by far the largest of them, with $I_{2j}$ coming second one or several orders of magnitude smaller, and the remaining two far behind the second one; this is perfectly confirmed by their values given in Table 6. Moreover, using the mean value theorem for integrals, one finds that there exists a radius ${r_j}$ such that
\be
\epsilon_j={I_{2j}}/{I_{1j}}=\Omega({r_j})\approx {r_j}^{-3/2}\ll 1,\qquad {r_j}\gg r_i>1\; .
\label{epssmall}
\ee 
The fact that the value $r_j$ is large enough follows from the radial behavior of the integrands; so, the number $\epsilon_j$ should be small, and indeed 
Table 6 gives it at $\sim 10^{-6}$, $\sim 10^{-3}$, and $\lesssim 10^{-1}$, 
for $r_o = 10^{-4}$, $100$, and $20$, respectively.
 
\subsubsection{Eigenfrequency dependence on the mode numbers $m,\;j,\;n$}

The above insights become particularly helpful for the discussion of the behavior of $|\sigma_{mnj}^{\pm}|$ given by formulas (\ref{sigma0}), (\ref{sigmam}), as functions of the mode numbers $n,\;m$ and $j$, assuming that all the unperturbed disk parameters, starting with $a$, are fixed by some thin disk model, such as the one described in the next section. 

The first relevant property is rather straightforward: apparently, all eigenfrequencies grow when $n$ increases (and the two other mode numbers stay fixed in the allowed range, as assumed in all other cases as well), as can be
seen in Tables 7 and 8.
The dependence on $m$ is more tricky, although it is not very difficult to see that $|\sigma_{mnj}^{+}|$ always decreases when $m(>0)$ increases. As for $|\sigma_{mnj}^{-}|$, it evidently grows with $m$ if $\epsilon_j^2- {I_{4j}}/{I_{1j}}\geq0$. It also grows in the opposite case when 
\[
m^2<\left(\pi n-I_{1j}\right)\left[2I_{1j}\left({I_{4j}}/{I_{1j}}-\epsilon_j^2\right)\right]^{-1}\; ,
\]
but starts decreasing when $m^2$ becomes larger than the number on the right of this inequality.

The dependence on the vertical mode number seems to be the most complicated until one notices that $\epsilon_j={I_{2j}}/{I_{1j}}$ would be entirely independent of $j$ if the parameter $g=1/(\Gamma-1)$ were constant along the disk radius. If this is the case, $K_j^g$ would also be constant, would be pulled outside the integrals, and would thus vanish from their ratio; see formulas (\ref{epsI12}). Hence the dependence of $\epsilon_j$ on $j$ is very weak at most, the same is true for the ratio ${I_{4j}}/{I_{1j}}$, and the trend is defined mostly by the ratio ${I_{3j}}/{I_{1j}}$, which appears to be roughly quadratic in $j$. 
The data of Tables 7 and 8 do not, however, confirm or disprove this observation, because the eigenfrequencies with different values of $j$ in them have also very different values of $n$.

Finally, one can notice that, by the definition (\ref{Kjg}),
\[
K_{j-1}^g\;({\rm even\; mode})=K_j^g\;({\rm odd\; mode})\; .
\]
Since the approximations (\ref{sigma0}), (\ref{sigmam}), and, in fact, all the eigenfrequencies found from the radial and vertical WKB equations, depend on $j$ through the combination $K_j^g$ only, the same equality is true for them as well:
\be
\sigma_{mnj-1}^{\pm}\;({\rm even\; mode}) = \sigma_{mnj}^{\pm}\;({\rm odd\; mode})\; ,
\label{sigmaident}
\ee
provided both sides of it exist for a given set $m,\; n,\;j$. Thus within the WKB approximation, two modes can be excited at each eigenfrequency; one of them is even, the other is odd.

\section{Unperturbed Hydrodynamic Disk Model}

To carry out the 
hydrodynamical
analysis of the accretion disk oscillations, one needs, in the first place, to formulate the 
assumptions about the unperturbed state. In addition to what has already 
been said in Section 2, we make all the relevant statements on the unperturbed disk model in this section. The results discussed in it were obtained by assuming 
that the opacity is mainly due to (optically thick) electron scattering. This 
condition holds for $r\lesssim 4\times 10^3(L/L_{Edd})^{2/3}$ \citep{st}, so our results at larger radii are only approximate.
($L_{Edd}\propto M$ is the Eddington luminosity.)

It is notable that the interior structure of the (zero buoyancy) accretion disk
enters our hydrodynamical formulation only through the index $\Gamma$ and the function
$\alpha(r)$, defined by equation (\ref{alpha}).
(This is not to be confused with the usual viscosity parameter, here 
denoted by $\alpha_*$).
The relativistic factor
\be
\beta\sqrt{g_{rr}} =  (1+a r^{-3/2}) [(1-3r^{-1}+2a r^{-3/2})
   (1 - 2 r^{-1} + a^2 r^{-2})]^{-1/2} \label{ss1}
\ee
that appears in the definition of $\alpha(r)$ is of order unity except for 
values of $a$ close to 1 and simultaneously $r$ close to $r_i \approx 1$. 
The value of the coefficient (\ref{ss1}) at $r=r_i(a)$ as function of $a$ 
is shown in Figure 2.

The values for the speed of sound are obtained from the fully relativistic 
results of \citet{nt}:
\begin{eqnarray}
c_s/c & = & 5.0 \times 10^{-3}
(L/L_{Edd})^{1/5} (\alphav M/M_\sun)^{-1/10}
\eta^{-1/5}(a)
r^{-9/20}
{\cal B}^{-1/5}(a,r) {\cal D}^{-1/10}(a,r)  \nonumber \\
  & & [{\cal Q}(a,r) + \delta {\cal Q}(a,r)]^{1/5} \qquad
{\rm (gas \ pressure \ dominance, \ }\Gamma = 5/3) \; , \label{ss53} \\
c_s/c & = & 1.3 \times
(L/L_{Edd})
\eta^{-1}(a)
r^{-3/2}
{\cal A}(a,r) {\cal B}^{-2}(a,r) {\cal D}^{-1/2}(a,r) {\cal E}^{-1/2}(a,r)
\nonumber \\
  & & [{\cal Q}(a,r) + \delta {\cal Q}(a,r)]  \qquad
{\rm (radiation \ pressure \ dominance, \ }\Gamma = 4/3) \; ,
\end{eqnarray}
where
\be
\delta {\cal Q}(a,r) \equiv (1 - F_*)
{\cal B }(r_i,a)
{\cal C}^{1/2}(r_i,a)
{\cal F}(r,a)
{\cal B}^{-1}(r,a)
(1 - 2 r^{-1} + a^2 r^{-2})^{1/2} \; . \label{ss4}
\ee
In these expressions, $\eta$ (shown in Figure 3) is the `efficiency factor' 
which relates the mass accretion rate to the luminosity: $L = \eta 
\dot{M}c^2$. (It is the energy lost by a unit of rest-mass energy as it accretes from far away to the inner edge of the disk.)
The constant $F_*$ is the fraction of the inflowing specific angular 
momentum at $r_i$ that is absorbed by the black hole (so there is a torque at $r_i$ unless 
$F_* = 1$).
The functions ${\cal A}$, ${\cal B}$, ${\cal D}$, and ${\cal E}$ are 
relativistic correction factors defined in \citet{nt}, which approach unity 
at large $r$, while ${\cal Q}$ [defined in \citet{nt,pt}] is the factor, 
proportional to the stress, which is responsible for making
$\alpha(r)$ singular at $r_i$ when there is no torque present at the inner 
edge. In this paper we consider the case in which there is such a torque, 
and thus $\delta {\cal Q}$ has been added to ${\cal Q}$ [see \cite{p}]. 
Expressions (\ref{ss1})--(\ref{ss4}) completely specify the key function 
$\alpha(r)$, defined by equation (\ref{alpha}), in the domains of both gas 
and radiation pressure dominance discussed next. In particular, the two critical powers involved in it and used in calculations are $\mu=0$ and $\nu=9/20$.

Since we deal with all kinds of modes in this paper, we need to consider the 
dependence of the adiabatic index $\Gamma$ on the radius \citep{ort2}. The regions where 
gas or radiation pressure dominates were obtained by making use of the 
solutions of \citet{nt}. We define the transition radius as the radius at 
which the gas pressure equals the radiation pressure. The results of these 
considerations are shown in Table 3. Gas pressure dominates for $r_i<r<R_1$ 
and $r>R_2$, while radiation pressure dominates for $R_1<r<R_2$. In our 
calculations we use $\Gamma = 5/3$ and $\Gamma = 4/3$ for these two respective regions.

Typical values for the disk density and central object mass make the 
neglect of self-gravity a common assumption. An order-of-magnitude analysis 
\citep{ort2} shows that self-gravitational
effects can be ignored, at both the unperturbed or the perturbed level 
(Cowling approximation), whenever (now in normal units)
\begin{equation}\label{sg-per}
  \frac{r}{GM/c^2} \ll \frac{10^{11}}{(M/M_{\odot})^{0.96}} \; .
\end{equation}
Even though this estimate assumes $\alphav = 0.1$ and $L = L_{Edd}$, 
condition (\ref{sg-per})
is only slightly less restrictive for lower values of the luminosity.
Thus, for stellar mass black holes the condition is satisfied by many 
orders of magnitude, while one needs to be more careful with supermassive 
black holes when studying the outer modes. A black hole mass of $10^8 
M_\sun$, for example, implies the constraint $r \ll 2\times 10^3 GM/c^2$.

\section{Numerical Results for Full--Disk P--Modes}

At this point, we calculate some eigenfrequencies, aiming at those with the lowest magnitudes consistent with the appropriate restrictions discussed below. We compute them for different values of $a$, $j$, $m$, $r_o$, considering only the even modes [see relation (\ref{sigmaident})]. We also consider full--disk p--modes with negative $\Psi$, essentially the only p--modes to be captured by our approach.

We make use of the unperturbed disk model described in the previous section, and choose the following (`typical') values for the parameters: $L/L_{Edd} = 0.1$, $\alphav = 0.1$, $M = 10 M_\sun$, and $F_* = 0.95$. We also choose $r_o=10^4$, typical of stellar mass black holes in low--mass X--ray binaries (LMXBs). In addition, we perform calculations corresponding to values of $r_o = 100$ and 20. Recall also that, as shown in Section 7.2, our results are robust regarding the boundary conditions at the disk edges.

Using the fact that $1 -\Psi/\omega_*^2\sim \omega_*^2$ [equation (\ref{univas}) and Figure 1] and that $|\sigma|/\Omega_{\perp}(r)\gg 1$ over most of the disk, one obtains the estimate
\be
r/\lambda_r(r)\sim 10^2\sigma^2r^3
\label{est}
\ee 
at radii $r\gg r_i$ from equation (\ref{lamr}), using also equation (\ref{ss53}). Since the integrand in equation (\ref{5.11}) is $1/\lambda_r(r)$, the integral is dominated by its value near $r_o$.

Note that the fundamental requirement allowing for our WKB separation of variables, $r/\lambda_r(r)\gg 1$, is thus enforced by its value near $r_i$. From equations (\ref{5.11}) and (\ref{est}), we then obtain the radial WKB requirement
\be
n\gg (r_o/r_i)^3 \; . \label{nlb}
\ee
This lower limit is seen to agree approximately with the minimum values of $n$ found in Tables 7 and 8, obtained from another WKB requirement discussed below.

We turn now to the issue of the restrictions on the eigenfrequencies to be found.
The first of them is 
\be
1 - \Psi/\omega_*^2 > 1\;
\label{keycond}
\ee
(WKB parameter `not small'), and our computations prove that this is the key limitation:  all other conditions are satisfied if this one holds.

Indeed, the restriction (\ref{keycond}), in view of relation (\ref{univas}), implies that
\be
|\omega_*| > \sqrt{0.75 K^g_j} \; . \label{cond-1}
\ee
The relevant values of $K^g_j$ for several lowest values of $j$ are given  in Table 5.
The mode existence criterion (\ref{condition-omega}),
\[
|\omega_*| > \pi(j + 1 + g/2)/\sqrt{2g} \; ,
\]
turns out to be valid whenever condition (\ref{cond-1}) is true, as one concludes from Table 4.

We begin the computations by calculating $I_{1j}, I_{2j}, I_{3j}, I_{4j},$ and
$\epsilon_j$ for different values of $j$ and $a$ (Table 6). The value $\epsilon = 2.04\times 10^{-6}$ is the same for all the $r_o = 10^4$ entries in Table 6;
for $r_o = 100$, it is $0.00171 < \epsilon < 0.00181$, while
for $r_o = 20$, one finds $0.0283 < \epsilon < 0.0325$. 

Tables 7 and 8 list the main results.
In them, $\sigma$ is calculated taking into account the dominant condition
(\ref{cond-1}), which determines the minimum radial mode number $n$ (and thus the lowest possible $|\sigma|$) according to the formulas (\ref{sigma0}) or (\ref{sigmam}). Namely, these lowest values are achieved when condition (\ref{cond-1}) is valid all the way through the disk except just a single point $r=r_c$ where the inequality (\ref{cond-1}) turns into an equality; lower values of $n$ (and $|\sigma|$) would give an interval of radii where (\ref{cond-1}) is violated. The critical radius $r_c$ is also listed in the tables; usually it coincides with the inner radius $r_i$, or the transition radius $R_1$, at which point the pressure becomes radiation dominated (see Section 8), and $K^g_j$ has a jump. In a few cases, the critical radius lies within the radiation-pressure dominated region; in such instances, $r_c$ is indicated in the table by the appropriate factor multiplying $R_1$.

It is important to mention that, as stated above, we might have missed eigenfrequencies with moderate magnitudes (corresponding to $|\omega_*|\sim 1$) in our calculations. Those are the eigenfrequencies whose valus are too high to be captured by the low frequency analysis of RD3, and at the same time too low to be properly described by the current WKB approach. Those missing should include some inner and outer modes, as well as full--disk modes with radial mode numbers somewhat smaller than the ones found here. 

For instance, for $m=0$ and $j=-1$ (and gas pressure dominance) the WKB criterion (\ref{cond-1}) gives $|\sigma|>(3/2)\Omega_\perp$. Now for most values of $a$, the maximum value of $\Omega_\perp(r)\gtrsim 3\kappa_{max}$. As also seen from Tables 1 and 7, this WKB criterion then gives $|\sigma|\gtrsim 5\kappa_{max}$, which should allow lower eigenfrequencies that are greater than $\kappa_{max}$ to exist. The ($m=0$) traveling acoustic waves found by \citet{hmk} and \citet{mt96,mt97} had a narrow range of frequencies near $\kappa_{max}$.

\section{Discussion}

As can be appreciated from the tables, for the 
hydrodynamical
modes that we found the usual trend is to have an eigenfrequency which increases monotonically with $a$, except for a possible maximum for high values of $a$ (like the low-frequency fundamental p--mode). The dimensional frequency $f = 3.23\times 10^4(M_\sun/M)|\sigma|$ Hz. For the mass of the central object that we adopted ($M = 10 M_\sun$), the minimum eigenfrequencies in Tables 7 and 8 are thus (in order) 378, 552, 568, 769, and 988 Hz; all but one corresponding to $a=0$. Moreover, the eigenfrequencies increase monotonically with $j$.

Comparing results for different values of the outer radius $r_o$, we find that the smaller the disk, the smaller the value of $n$ for a given eigenfrequency. Also note that the value of the lowest eigenfrequency is independent of $r_o$. This is because the eigenfrequency is calculated using condition (\ref{cond-1}).

It has been found that the pair of high frequency QPOs seen in a few black hole LMXBs have a frequency ratio close to 3/2 \citep{ak,mr06}. We note that the ratio of the $j=1$ to the $j=0$ eigenfrequencies of the $m=0$ modes is 1.57, independent of the value of $a$, as seen in Table 7 (for $r_o=10^4$, but it is the same for the other values). They have roughly the same minimum radial numbers $n$. (For $m=1$, there are seen to be some with a ratio of 1.45.) However, these frequencies are part of an effective continuum of radial mode numbers which extends to higher (and possibly somehat lower) frequencies, as also discussed below. We do not know any obvious way in which such a ratio could be observed. 

Another possibility to explain the observed ratio is related to a pair of low--frequency fundamental ($m=j=0$) p--modes found in RD3. According to Table 1 of this paper, the ratio of the eigenfrequencies of such modes with radial mode numbers $n=2$ and $n=0$ is 1.53.  However, the capture zone of these modes seems too small to produce the observed amount of luminosity modulation. 

From this point of view, a better candidate is the $m=3$ and $m=2$ pair of g--modes. As found in RD1, their eigenfrequencies $|\sigma|\cong m\Omega(r_i)$ for $m>1$, and their radial and vertical mode splitting is much smaller. Of course, it is not obvious why the $m=0$ and $m=1$ g--modes, as well as the $j=-1$ p--mode (Table 7) and the $n=1$ low--frequency p--mode (RD3), are not also observed. However, the (turbulent?) excitation of modes is not well understood.      

We should note that \citet{abt} found no g--modes in their shearing--box MHD simulations of a limited radial region of an accretion disk. There were some indications of the generation of p--modes within the MRI--induced turbulence, however.

The full--disk p--modes studied in this paper share some characteristics with the other modes studied previously. For example, their eigenfrequencies show a dependence on the properties of the central object and are rather insensitive to changes in the luminosity of the disk.
Like most of the other modes, the properties of the full--disk p--modes are influenced by effects of general relativity in a strong field regime, and like the g--modes and some of the c--modes, they include the hottest part 
of the disk.

On the other hand, the full--disk p--modes have some distinct characteristics. In addition to the fact that they cover the whole disk, there seem to be no full--disk p--modes with low or moderate values of the radial mode number $n$. We also note that in the outer parts of the disk, their eigenfunctions mainly occupy only a small layer close to the surfaces of the accretion disk. 

The radial wavelength of a minimum $n$ mode is given by equation (\ref{est}) as $\lambda_r \sim (10\sigma r)^{-2}$. The observable effects of such modes are hard to perceive because their integrated luminosity fluctuations will average to effectively zero over the disk. It would seem that the most likely to be observed would be those axisymmetric modes whose frequency $|\sigma|$ lies closest to the peak value of $\kappa(r)$ (as was the case for the traveling waves mentioned above). In this case, the wavelength would be largest in that region, which is also located near the hottest part of the accretion disk.

We now note some important effects of viscosity on these (adiabatic) oscillations. 
If one applies the analysis developed in \citet{ort} to the full--disk $m=0$ p--modes, one finds that they are viscously unstable, with a growth rate $1/\tau \gtrsim \alpha_*\sigma^2 /\Omega(r_o)$ obtained approximately. (A similar result holds for the other modes with small values of $m$.) Here $\alpha_*$ is the ratio of the shear to the total pressure, with the shear viscosity $\eta\sim\alpha_* h^2\Omega$. This instability thus exists in both radiation and gas pressure dominated regions of the disk, unlike the axisymmetric Lightman--Eardley instability \citep{kfm}. Our approach generalizes the local analysis of acoustic waves, which are also found to be viscously unstable under certain conditions \citep{kfm}.

Although this result violates the assumption that the imaginary part of the eigenfrequency is small compared to the real part, it does suggest that the accretion disk is viscously unstable. This may well be a significant source of turbulence. In addition to the fact that the viscous width of the modes is much greater than the frequency separation of the radial harmonics, this growth would lead to the production of an effective continuum of excitation. We hope to investigate viscous damping and/or instability within a broader framework in the future, including a more realistic model of the effective (magnetically--generated) turbulent viscosity.

Various p--modes were studied here in rather general terms, employing a WKB approach to 
solve the eigenvalue  problem for the separated differential equations. 
As we have explained in Section 4, the nonexistence of some of these modes can be understood in physical terms taking into account the necessary continuity of the solutions together with the fact that formal conditions must be met over rather large radial regions containing changing physical 
conditions (namely, the transition from gas pressure to radiation pressure domination in the inner disk).

On the other hand, there exists the possibility that some other modes do exist but were not found due to the limitations of our method. Those limitations follow from the character of the WKB approach, only valid in 
principle for sufficiently small wavelengths. Applied to our study, this means that WKB is only strictly applicable when $|\omega_*| \gg 1$, although there are indications that it works for $|\omega_*|\sim 1$ as well. Having said this, it must be stressed that the modes found in these and the previous papers of the series provide a basis for dealing with the general problem of perturbations in thin accretion disks, assuming that the classes we have considered constitute a complete set.

\acknowledgments

This work was supported by NASA grant NAS 8-39225 to Gravity Probe B.
M. Ortega was supported by grant 075-2002 (Incentivos) of Ministerio
de Ciencia y Tecnolog\'{\i}a, Costa Rica, and by grant 829-A3-078 of
Vicerrector\'{\i}a de Investigaci\'on, Universidad de Costa Rica.
We are grateful to Lev Kapitanski for his help with proving the spectrum estimate (\ref{6.5}).

\section*{Appendix. Lower Bound for Vertical Eigenvalues}

We introduce a new unknown function in the vertical equation (\ref{eq:4}) as
$$
v(y)=(1-y^2)^{\delta/2}V_y(y),\qquad V_y(y)=(1-y^2)^{-\delta/2}v(y)\; ,
\eqno(A.1)
$$
with some real $\delta$. As $V_y(y)$ and its derivative are finite at 
$y^2=1$, for $v(y)$ and its derivative at $y^2\to1-0$ we obtain
$$
v(y)={\cal O}\left((1-y^2)^{\delta/2}\right),\qquad v^{'}(y)={\cal 
O}\left((1-y^2)^{\delta/2-1}\right)\; .
\eqno(A.2)
$$
Equation (\ref{eq:4}) in terms of $v(y)$ reads:
$$
(1-y^2)\,v^{''} + 
2y\,(g-\delta)\,v^{'}-a_1(1-y^2)\,v+a_2(1-y^2)^{-1}v+\lambda(1-y^2)\,v
= 0
\; ,
\eqno(A.3)
$$
where
$$
\lambda\equiv 2g\omega_*^2 \left(\Psi/\omega_*^2-1\right)
\; ,
\eqno(A.4)
$$
and
$$
a_1(\delta,g,\omega_*^2)\equiv\delta^2-(2g-1)\delta-2g\omega_*^2=a_2(\delta,g)-\delta-2g\omega_*^2,
$$
$$
\eqno(A.5)
$$
$$
a_2(\delta,g)\equiv\delta^2-2(g-1)\delta=(\delta-g+1)^2-(g-1)^2\; .
$$

To make the following formulas shorter, we introduce some notations. Given 
any real $\theta$ and any function $f(y)$ on $y\in(-1,\,1)$, we set a 
one-parameter family of norms
$$
||f||^2_\theta\equiv\int_{-1}^{1}(1-y^2)^\theta f^2(y)\,dy\; .
\eqno(A.6)
$$
Since, for an arbitrary $\epsilon>0$,
$$
(1-y^2)^\theta f^2=\left[(1-y^2)^{(\theta/2+1/2)} f\right]\,
\left[(1-y^2)^{(\theta/2-1/2)} f\right]\;\leq\;
\frac{\epsilon}{2}\,(1-y^2)^{(\theta+1)} 
f^2\;+\;\frac{1}{2\epsilon}\,(1-y^2)^{(\theta-1)} f^2\;,
$$
we have an important inequality
$$
||f||^2_\theta\;\leq\;
\frac{\epsilon}{2}\,||f||^2_{\theta+1}\;+\;\frac{1}{2\epsilon}\,||f||^2_{\theta-1}\;,
\eqno(A.7)
$$
provided that the last norm, and hence all of them, is finite (evidently, 
$||f||^2_{\theta}\leq||f||^2_{\theta^{'}}$ for $\theta>\theta^{'}$).

Choose now some real $\gamma$ satisfying
$$
\gamma+\delta>0\; ,
\eqno(A.8)
$$
multiply equation (A.3) by $(1-y^2)^{\gamma}v(y)$, and integrate over 
$(-1,\,1)$. Because of the above condition and formulas (A.2), all the 
integrals converge and all the double substitutions at $y=\pm1$ vanish in 
the following repeated integration by parts [we drop the limits $\pm1$ in 
all the integrals and use the identity $y^2=1-\left(1-y^2\right)$ wherever 
necessary]:
$$
\int(1-y^2)^{\gamma+1}vv^{''}\,dy=-||v^{'}||^2_{\gamma+1}+
2(\gamma+1)\int\,y(1-y^2)^{\gamma}vv^{'}\,dy\; ;
$$
$$
2\int\,y(1-y^2)^{\gamma}vv^{'}\,dy=-||v||^2_{\gamma}+
2\gamma\int\,y^2(1-y^2)^{\gamma-1}v^{2}\,dy=
$$
$$
-\,||v||^2_{\gamma}+2\gamma\,||v||^2_{\gamma-1}-2\gamma\,||v||^2_{\gamma}=
-(2\gamma+1)\,||v||^2_{\gamma}+2\gamma\,||v||^2_{\gamma-1}\; .
$$
With these results, the described integration of the equation (A.3) yields
$$
\lambda||v||^2_{\gamma+1}\,=\,
||v^{'}||^2_{\gamma+1}+p_1\,||v||^2_{\gamma}-p_2\,||v||^2_{\gamma-1}\,\geq\,
p_1\,||v||^2_{\gamma}-p_2\,||v||^2_{\gamma-1}\;
\eqno(A.9)
$$
for any eigenfunction $v$ corresponding to the eigenvalue $\lambda$ [they 
are related to the eigenfunction $V_y$ and eigenvalue $\Psi$ by equalities 
(A.1) and (A.4), respectively]. Here
$$
p_1=p_1(\gamma,\delta,g,\omega_*^2)=
(2\gamma+1)(\gamma+1+\delta-g)+a_1(\delta,g,\omega_*^2)\; ,
$$
$$
\eqno(A.10)
$$
$$
p_2=p_2(\gamma,\delta,g)=
2\gamma(\gamma+1+\delta-g)+a_2(\delta,g)\; .
$$

{}From the point of getting the lower bound for $\lambda$ the problem with 
the inequality (A.9) is that the norm on the left is smaller than 
both norms on the right. One way to resolve it seems to be the following.

Suppose that for some given $g>0,\, \omega_*^2\geq0$ there exist $\gamma$ 
and $\delta$ satisfying inequality (A.8) such that
$$
p_1(\gamma,\delta,g,\omega_*^2)<0,\qquad
p_2(\gamma,\delta,g)<0\; ,
\eqno(A.11)
$$
which makes the coefficient in front of the $\gamma$--norm in (A.9) 
negative, while the other one, in front of the biggest $(\gamma-1)$--norm, 
positive. Then we can use inequality (A.7) with $f=v,\,\theta=\gamma$ to 
write (recall that $p_1$ is negative),
$$
p_1||v||^2_\gamma\;\geq\;
\frac{p_1\epsilon}{2}\,||v||^2_{\gamma+1}\;+\;\frac{p_1}{2\epsilon}\,||v||^2_{\gamma-1}\;,
$$
and then set
$$
\epsilon=\frac{p_1}{2p_2}>0
$$
to eliminate the biggest norm in the inequality (A.9). The latter thus 
reduces to
$$
\lambda||v||^2_{\gamma+1}\;\geq\;
\frac{p_1^2}{4p_2}\,||v||^2_{\gamma+1}\; ,
$$
or simply to
$$
\lambda\;\geq\;\frac{1}{4}\,
\frac{p_1^2(\gamma,\delta,g,\omega_*^2)}{p_2(\gamma,\delta,g,)}\;
\eqno(A.12)
$$
[recall that $(p_1^2/p_2)<0$ by the assumption (A.9)].

This is the desired lower bound of the spectrum. If moreover $\Pi=\Pi(g, 
\omega_*^2)$ is the set of all admissible values $\{\gamma,\delta\}$ for 
given $g$ and $\omega_*^2$, then (A.12)  provides exactly the estimate 
(\ref{6.5}) with the best constant
$$
\Lambda(g,\omega_*^2)=-\frac{1}{4}\,
\sup_{\{\gamma,\delta\}\in\Pi}\; 
\frac{p_1^2(\gamma,\delta,g,\omega_*^2)}{p_2(\gamma,\delta,g,)}=
\frac{1}{4}\,
\inf_{\{\gamma,\delta\}\in\Pi}\; 
\frac{p_1^2(\gamma,\delta,g,\omega_*^2)}{|p_2(\gamma,\delta,g,)|}\geq 0\; .
\eqno(A.13)
$$

The only thing that remains to be demonstrated is that the set $\Pi(g, 
\omega_*^2)$ is not empty, in other words, that the inequalities (A.8) and 
(A.11) are always compatible for at least one pair $\{\gamma,\delta\}$. The 
help with solving those three inequalities, two of which are quadratic in 
both $\gamma$ and $\delta$, comes from a geometrical approach which turns 
out effective if we reparametrize the inequalities using
$$
x=\gamma+\delta +1-g
\eqno(A.14)
$$
instead of $\delta$. It is easy to see that expressions (A.10) for $p_1$ 
and $p_2$ can be now rewritten as
$$
p_1=(\gamma+1/2)^2+x^2-r_{I}^2,\qquad
p_2=\gamma^2+x^2-r_{II}^2\; ,
\eqno(A.15)
$$
with
$$
r_I=r_I(g,\omega_*^2)=\sqrt{(g-1/2)^2+2g\omega_*^2},\qquad 
r_{II}=r_{II}(g)=|g-1|\; .
\eqno(A.16)
$$
Therefore inequalities (A.11) specify two open circles, $C_{I}$ and 
$C_{II}$, in the plane $\{\gamma,x\}$, while condition (A.8) corresponds to 
the open half--plane $H$ specified by $x>1-g$. Everything depends now on 
how $g$ compares to unity.

Indeed, if $g>1\, (1<\Gamma<2)$, then
$$
r_I=\sqrt{(g-1/2)^2+2g\omega_*^2}>r_{II}+1/2=g-1+1/2=g-1/2,
$$
so that the circle $C_{II}$ lies completely in the circle $C_{I}$. It also 
belongs to the half--plane $H$, hence, in fact, this circle is the 
intersection of all the three domains. Therefore we have the exact lower 
bound of the eigenvalues (A.13) with
$$
\Pi=\Pi(g)=C_{II}(g)=
\{\{\gamma,\delta\}\,:\, \gamma^2+(\gamma+\delta +1-g)^2<(g-1)^2\}
=\{\{\gamma,x\}\,:\, \gamma^2+x^2<r_{II}^2\}\; .
$$
It is now straightforward to see that the minimum value in (A.13) is 
achieved when $x=0$, so that
$$
\Lambda(g,\omega_*^2)=
\frac{1}{4}\,
\inf_{|\gamma|<r_{II}}\; 
\frac{(\gamma^2+\gamma+1/4-r_{I}^2)^2}{r_{II}^2-\gamma^2}> 0\; .
\eqno(A.17)
$$
However, finding explicit value of $\Lambda$ requires solving analytically 
a cubic equation with the parameters $r_{I}$ and $r_{II}$ in the 
coefficients, which makes it impractical. On the other hand, any particular 
values of parameters from $C_{II}$ can be introduced into (A.12) to give 
the estimate with a somewhat worse but explicit constant. In particular, 
inequality (\ref{6.6}) is obtained when $x=0\,(\delta=g-1)$ and $\gamma=0$, 
which corresponds to the center of the circle $C_{II}$.

If, however, $g=1\, (\Gamma=2)$, then the circle $C_{II}$ shrinks into a 
point, $p_2$ is nonnegative everywhere, and no estimate is available. 
Moreover, for $g<1\, (\Gamma>2)$, the circle $C_{II}$ is outside $H$, 
inequalities (A.8) and
(A.11) are incompatible, $\Pi$ is empty, and no lower bound for the 
vertical eigenvalue spectrum can be gotten in this way. Thus the question 
whether such a bound exists at all when $g<1\, (\Gamma>2)$ remains so far 
unanswered.

One could note that $\lambda$ would be nonnegative (no p--modes would 
exist) if $p_1\geq0$ and $p_2\leq0$ in the inequality (A.9). The above 
analysis shows, nevertheless, that such situation is impossible for any $g>0$.

\begin{deluxetable}{ccccc}
\tablecolumns{5}
\tablewidth{0pc}
\tablecaption{Eigenfrequency boundary $-\sigma_-=\max[\kappa(r) + m 
\Omega(r)]$;
$\Omega(r_i)$ (last row)}
\tablehead{
\colhead{} &
\multicolumn{4}{c}{$a$} \\
\cline{2-5} \\
\colhead{$m$}  &
\colhead{$0$}    &
\colhead{$0.5$}    &
\colhead{$a_\perp \approx 0.953$}    &
\colhead{$0.998$}}
\startdata
0&0.02210&0.03312&0.06597&0.07492 \\
1&0.07720&0.1209\phn&0.2962\phn&0.4277\phn \\
2&0.1414\phn&0.2244\phn&0.5658\phn&0.8457\phn \\ \hline
 $\Omega(r_i)$ &0.06804   &0.1086\phn&0.2784\phn&0.4213\phn
\enddata
\end{deluxetable}

\begin{deluxetable}{ccccc}
\tablecolumns{5}
\tablewidth{0pc}
\tablecaption{Radii that maximize $\kappa(r) + m \Omega(r)$}
\tablehead{
\colhead{} &
\multicolumn{4}{c}{$a$} \\
\cline{2-5} \\
\colhead{$m$}  &
\colhead{$0$}    &
\colhead{$0.5$}    &
\colhead{$a_\perp$}    &
\colhead{$0.998$}}
\startdata
0& 8.0000 & 5.7628 & 2.9639 & 2.1929 \\
1& 6.4000 & 4.6377 & 2.0073 & 1.2608 \\
2& 6.1407 & 4.3000 & 1.9497 & 1.2424 \\ 
\enddata
\end{deluxetable}

\begin{deluxetable}{cccccc}
\tablecolumns{4}
\tablewidth{0pc}
\tablecaption{Relevant radii in the Novikov \& Thorne disk solution, for
$M/M_{\odot} = 10$ and $L/L_{Edd} = 0.1$}
\tablehead{
\colhead{$a$} &
\colhead{$r_i$} &
\colhead{$R_1$} &
\colhead{$R_2$} &
   }
\startdata
$0\phantom{.000}$ & $6\phantom{.00000}$ & $6.92$ & $140$  \\
$0.5\phantom{00}$ & $4.233\phn\phn$ & $4.81$ & $100$ \\
$0.953$           & $1.9099\phn$ & $2.06$ & $\phn 53$ \\
$0.998$           & $1.23697$ & $1.26$ & $\phn 34$ \\
\enddata
\end{deluxetable}

\begin{deluxetable}{ccc}
\tablecolumns{3}
\tablewidth{0pc}
\tablecaption{Values of $\pi(j + 1 + g/2)/\sqrt{2g}$ for even modes 
with $g=3/2$ and $g=3$ (gas and radiation-pressure dominated regions, respectivelly), relevant for
condition (\ref{condition-omega}).}
\tablehead{
\colhead{$j$} &
\colhead{  }  &
\colhead{  }
  }
\startdata
$-1$ & $1.36$ & $1.92$ \\
$0$ & $3.17$ & $3.21$ \\
$1$ & $4.99$ & $4.49$ \\
$2$ & $6.80$ & $5.77$ \\
$3$ & $8.62$ & $7.05$ \\
\enddata
\end{deluxetable}

\begin{deluxetable}{ccc}
\tablecolumns{3}
\tablewidth{0pc}
\tablecaption{Values of $K^g_j$ for even modes with $g=3/2$ and $g=3$ (gas-
and radiation-pressure dominated regions, respectivelly), 
relevant for the condition $1 - \Psi/\omega_*^2 > 1$.}
\tablehead{
\colhead{$j$} &
\colhead{$K^{3/2}_j$}  &
\colhead{$K^{3}_j$}
  }
\startdata
$-1$ & \phantom{11}3\phantom{.33}  & \phantom{1}6\phantom{.33} \\
$0$ & \phantom{1}16.33  & 16.67 \\
$1$ & \phantom{1}40.33  & 32.67 \\
$2$ & \phantom{1}75  \phantom{.33}  &  54\phantom{.33}\\
$3$ & 120.33 & 80.67 \\
\enddata
\end{deluxetable}

\begin{deluxetable}{cccccc}
\tablecolumns{6}
\tablewidth{0pc}
\tablecaption{Values of $I_{i}$ (shorthand
for $I_{ij}$).} 
\noindent
\tablehead{
\colhead{} &
\colhead{} &
\colhead{} &
\multicolumn{3}{c}{$a$} \\
\cline{4-6} \\
\colhead{$r_o$} &
\colhead{$j$} &
\colhead{} &
\colhead{$0$}    &
\colhead{$0.5$}    &
\colhead{$a_\perp$} 
  }
\startdata
$20$ &
$-1$   & $I_1$
&  $ 4.58 \,{10}^{4}$   
&  $ 4.36 \,{10}^{4}$
&  $ 6.36 \,{10}^{4}$  
\\
 &
$ $   & $I_2$
&  $ 1.40 \,{10}^{3}$   
&  $ 1.42 \,{10}^{3}$
&  $ 2.27 \,{10}^{3}$
\\
 &
   & $I_3$
&  $ 24.5 $  
&  $ 30.2 $
&  $ 55.9 $
\\
 &
   & $I_4$
&  $ 60.6 $  
&  $ 85.8 $
&  $ 310 $  
\\
\cline{2-6}
 &
$0$   & $I_1$
&   $ 2.56 \,{10}^{4}$  
&   $ 2.51 \,{10}^{4}$
&   $ 3.77 \,{10}^{4}$
\\
 &
$ $   & $I_2$
&  $ 723 $   
&  $ 754 $
&  $ 1.22 \,{10}^{3}$
\\
 &
   & $I_3$
&  $ 57.1  $  
&  $ 70.0  $
&  $ 125  $
\\
 &
   & $I_4$
&  $ 29.2 $  
&  $ 41.6 $
&  $ 149  $
\\
\hline 
$100$ &
$-1$   & $I_1$
&  $ 7.32 \,{10}^{6}$   
&  $ 9.75 \,{10}^{6}$
&  $ 3.48 \,{10}^{7}$  
\\
 &
$ $   & $I_2$
&  $ 1.35 \,{10}^{4}$   
&  $ 1.73 \,{10}^{4}$
&  $ 5.71 \,{10}^{4}$
\\
 &
   & $I_3$
&  $ 33.2 $  
&  $ 40.9 $
&  $ 59.4 $
\\
 &
   & $I_4$
&  $ 88.0 $  
&  $ 121 $
&  $ 417 $  
\\
\cline{2-6}
 &
$0$   & $I_1$
&   $ 4.39 \,{10}^{6}$  
&   $ 5.85 \,{10}^{6}$
&   $ 1.52 \,{10}^{7}$
\\
 &
$ $   & $I_2$
&  $ 7.97 \,{10}^{3}$   
&  $ 1.03 \,{10}^{4}$
&  $ 2.60 \,{10}^{4}$
\\
 &
   & $I_3$
&  $ 84.2 $  
&  $ 104 $
&  $ 207 $
\\
 &
   & $I_4$
&  $ 45.6 $  
&  $ 62.8 $
&  $ 201 $
\\ 
\hline
$10^4$ &
$-1$   & $I_1$
&  $2.21 \,{10}^{13}$   
&  $2.38\,{10}^{13}$
&  $2.82\,{10}^{13}$  
\\
 &
$ $   & $I_2$
&  $4.51 \,{10}^{7\phantom{0}}$   
&  $4.85 \,{10}^{7\phantom{0}}$
&  $5.75 \,{10}^{7\phantom{0}}$
\\
 &
   & $I_3$
&  $-2.04$  
&  $-3.19$
&  $7.05 $
\\
 &
   & $I_4$
&  $ 417 $  
&  $ 494 $
&  $ 857 $  
\\
\cline{2-6}
 &
$0$   & $I_1$
&   $9.49 \,{10}^{12}$  
&   $1.02\,{10}^{13}$
&   $1.21\,{10}^{13}$
\\
 &
$ $   & $I_2$
&  $1.93 \,{10}^{7\phantom{0}}$   
&  $2.08 \,{10}^{7\phantom{0}}$
&  $2.47 \,{10}^{7\phantom{0}}$
\\
 &
   & $I_3$
&  $ 305 $  
&  $ 352 $
&  $ 498 $
\\
 &
   & $I_4$
&  $ 188 $  
&  $ 223 $
&  $ 390 $
\\ 
\enddata
\end{deluxetable}

\begin{deluxetable}{cccccc}
\tablecolumns{6}
\tablewidth{0pc}
\tablecaption{Lowest values of the eigenfrequency $|\sigma|$, 
the radial number 
$n$ and the critical radius $r_c$ for
full--disk modes with $\Psi<0$ and $m = 0$.}
\noindent
\tablehead{
\colhead{} &
\colhead{} &
\colhead{} &
\multicolumn{3}{c}{$a$} \\
\cline{4-6} \\
\colhead{$r_o$} &
\colhead{$j$} &
\colhead{} &
\colhead{$0$}    &
\colhead{$0.5$}    &
\colhead{$a_\perp$} 
  }
\startdata
 $20$ &
   $-1$& $|\sigma|$  
& $0.117$  
&  $0.176$  
&  $0.322$  
\\
 &
   & $n$   
&  $205 $  
&   $439$      
&   $2.12 \,{10}^{3}$      
\\
 &   
 &  $r_c$
&  $R_1$
&  $R_1$
&  $R_1$ 
\\
\cline{2-6}
 &
   $0$&  $|\sigma|$ 
&   $ 0.238 $  
&    $0.343 $   
&    $0.537 $  
\\
 &
   & $n$   
&  $479 $
&  $960 $  
&  $3.50 \,{10}^{3} $
\\
 &
 &  $r_c$
&  $r_i$ 
&  $r_i$
&  $R_1$
\\ 
\hline
$100$ &
   $-1$& $|\sigma|$  
& $0.117$  
&  $0.176$  
&  $0.322$  
\\
 &
   & $n$   
&  $3.17 \,{10}^{4}$  
&   $9.65 \,{10}^{4}$      
&   $1.15 \,{10}^{6}$      
\\
 &   
 &  $r_c$
&  $R_1$
&  $R_1$
&  $R_1$ 
\\
\cline{2-6}
 &
   $0$&  $|\sigma|$ 
&   $ 0.238 $  
&    $0.343 $   
&    $0.537 $  
\\
 &
   & $n$   
&  $7.93 \,{10}^{4} $
&  $2.19 \,{10}^{5}$  
&  $1.40 \,{10}^{6} $
\\
 &
 &  $r_c$
&  $r_i$ 
&  $r_i$
&  $R_1$
\\ 
\hline
$10^4$ &
   $-1$& $|\sigma|$  
& $0.117$  
&  $0.176$  
&  $0.322$  
\\
 &
   & $n$   
&  $9.57 \,{10}^{10}$  
&   $2.35 \,{10}^{11}$      
&   $ 9.33 \,{10}^{11}$      
\\
 &   
 &  $r_c$
&  $R_1$
&  $R_1$
&  $R_1$ 
\\
\cline{2-6}
 &
   $0$&  $|\sigma|$ 
&   $ 0.238 $  
&    $0.343 $   
&    $0.537 $  
\\
 &
   & $n$   
&  $ 1.71 \,{10}^{11} $
&  $3.81 \,{10}^{11}$  
&  $1.11 \,{10}^{12} $
\\
 &
 &  $r_c$
&  $r_i$ 
&  $r_i$
&  $R_1$
\\
\cline{2-6}
 &
   $1$&  $|\sigma|$ 
&   $ 0.374 $  
&    $0.538$   
&    $0.842$  
\\
 &
   & $n$   
&  $ 2.69 \,{10}^{11} $
&  $5.98 \,{10}^{11}$  
&  $1.74 \,{10}^{12} $
\\
 &
 &  $r_c$
&  $r_i$ 
&  $r_i$
&  $r_i$
\\
\enddata
\end{deluxetable}

\begin{deluxetable}{cccccc}
\tablecolumns{6}
\tablewidth{0pc}
\tablecaption{Lowest values of the eigenfrequency $|\sigma|$, the radial 
number $n$ and the critical radius $r_c$ for
full--disk modes; $\Psi<0, \,\,m = 1$, $\sigma < 0$ case.}
\noindent
\tablehead{
\colhead{} &
\colhead{} &
\colhead{} &
\multicolumn{3}{c}{$a$} \\
\cline{4-6} \\
\colhead{$r_o$} &
\colhead{$j$} &
\colhead{} &
\colhead{$0$}    &
\colhead{$0.5$}    &
\colhead{$a_\perp$} 
  }
\startdata
 $20$ &
   $-1$& $|\sigma|$  
& $0.171$  
&  $0.267$
&  $0.578$  
\\
 &
   & $n$   
&  $176$  
&   $827$      
&   $7.02 \, {10}^{3}$      
\\
 &   
 &  $r_c$
&  $R_1$
&  $R_1$
&  $R_1$ 
\\
\cline{2-6}
 &
   $0$&  $|\sigma|$ 
&   $ 0.306 $  
&    $0.451 $   
&    $0.814 $  
\\
 &
   & $n$   
&  $ 707 $
&  $ 1.65 \,{10}^{3}$  
&  $ 8.30 \,{10}^{3} $
\\
 &
 &  $r_c$
&  $r_i$ 
&  $r_i$
&  $r_i$
\\ 
\hline
$100$ &
   $-1$& $|\sigma|$  
& $0.171$  
&  $0.267$
&  $0.578$  
\\
 &
   & $n$   
&   $6.57 \,{10}^{4}$  
&   $2.18 \,{10}^{5}$      
&   $3.70 \,{10}^{6}$      
\\
 &   
 &  $r_c$
&  $R_1$
&  $R_1$
&  $R_1$ 
\\
\cline{2-6}
 &
   $0$&  $|\sigma|$ 
&   $ 0.306 $  
&    $0.451 $   
&    $0.814 $  
\\
 &
   & $n$   
&  $1.30 \,{10}^{5} $
&  $3.79 \,{10}^{5}$  
&  $3.22 \,{10}^{6} $
\\
 &
 &  $r_c$
&  $r_i$ 
&  $r_i$
&  $r_i$
\\ 
\hline
$10^4$ &
   $-1$& $|\sigma|$  
& $0.171$  
&  $0.267$
&  $0.578$  
\\
 &
   & $n$   
&  $2.06 \,{10}^{11}$  
&   $5.40 \,{10}^{11}$      
&   $3.00 \,{10}^{12}$      
\\
 &   
 &  $r_c$
&  $R_1$
&  $R_1$
&  $R_1$ 
\\
\cline{2-6}
 &
   $0$&  $|\sigma|$ 
&   $ 0.306 $  
&    $0.451 $   
&    $0.814 $  
\\
 &
   & $n$   
&  $2.83 \,{10}^{11} $
&  $6.60 \,{10}^{11}$  
&  $2.55 \,{10}^{12} $
\\
 &
 &  $r_c$
&  $r_i$ 
&  $r_i$
&  $r_i$
\\
\cline{2-6}
 &
   $1$&  $|\sigma|$ 
&   $ 0.442 $  
&    $0.647$   
&    $1.12$  
\\
 &
   & $n$   
&  $3.76 \,{10}^{11} $
&  $8.65 \,{10}^{11}$  
&  $3.08 \,{10}^{12} $
\\
 &
 &  $r_c$
&  $r_i$ 
&  $r_i$
&  $r_i$
\\
\enddata
\end{deluxetable}

\newpage

\begin{figure}
\figurenum{1}
\epsscale{1}
\plotone{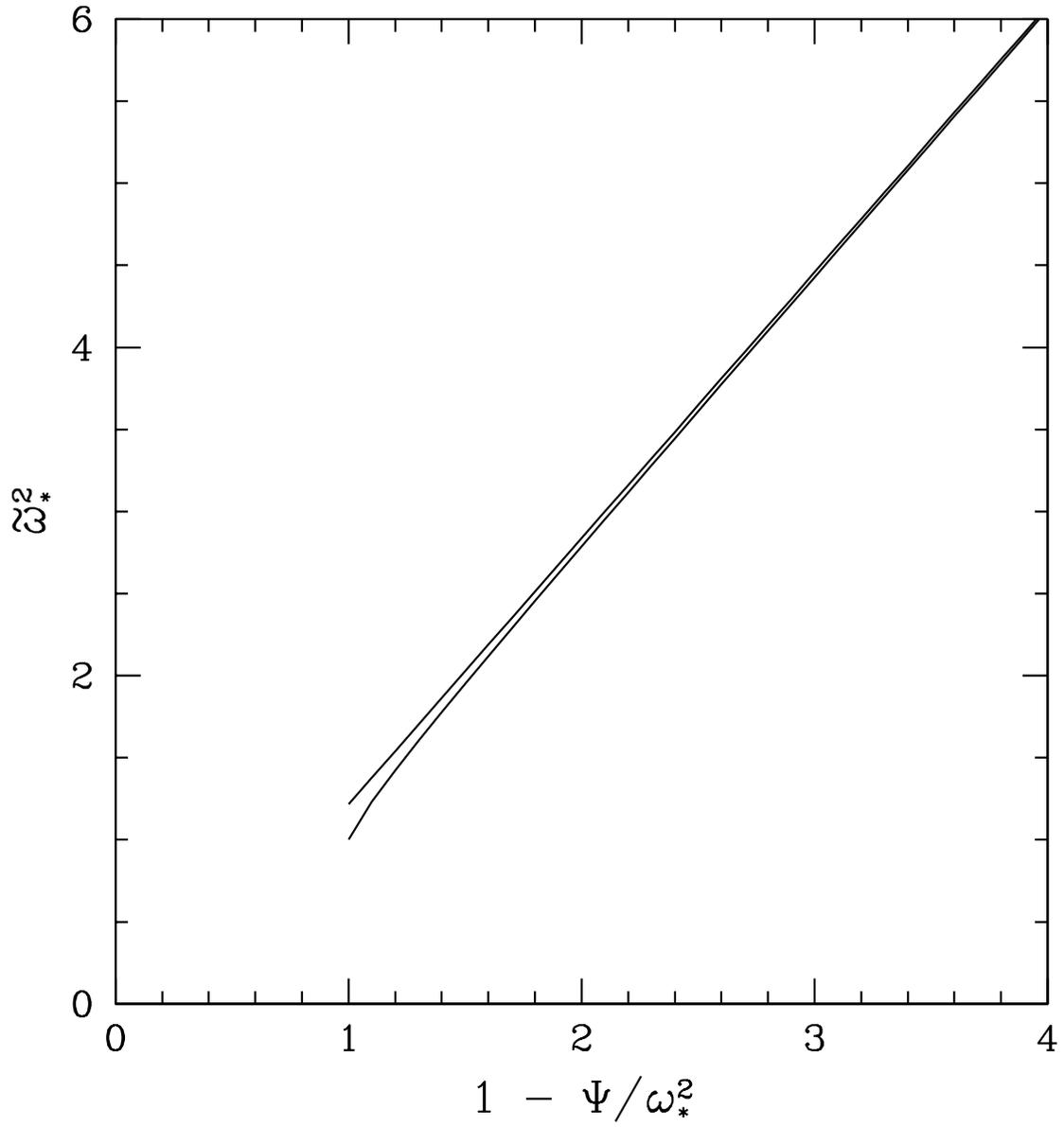}
\caption{Universal curve for the vertical WKB equation from 
the analytical asymptotic (top) and numerical (bottom) solutions.}
\end{figure}

\newpage

\begin{figure}
\figurenum{2}
\epsscale{1}
\plotone{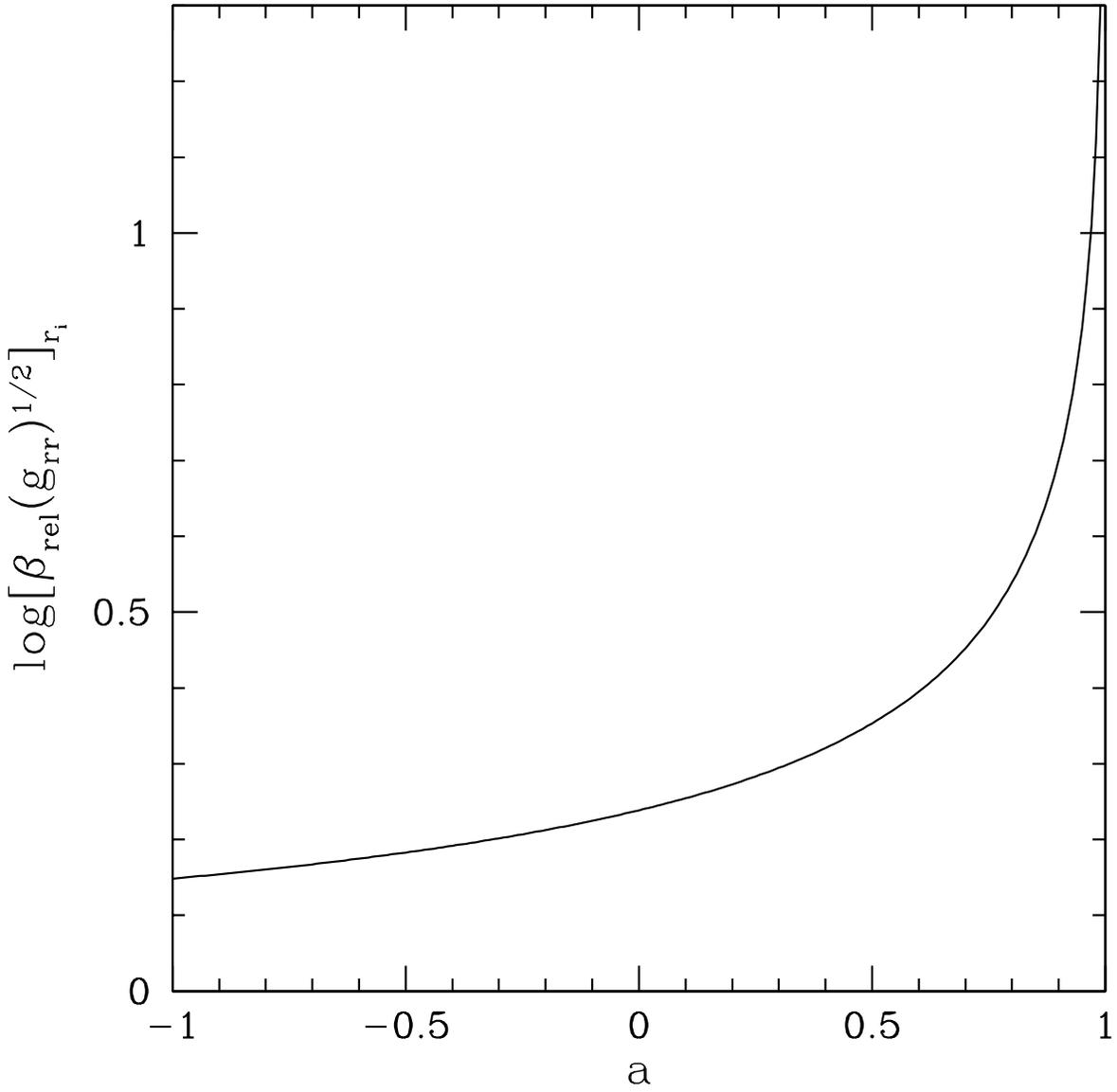}
\caption{Dependence of $(\beta\sqrt{g_{rr}})_{r_i}$ on the angular 
momentum of the black hole.}
\end{figure}

\newpage

\begin{figure}
\figurenum{3}
\epsscale{1}
\plotone{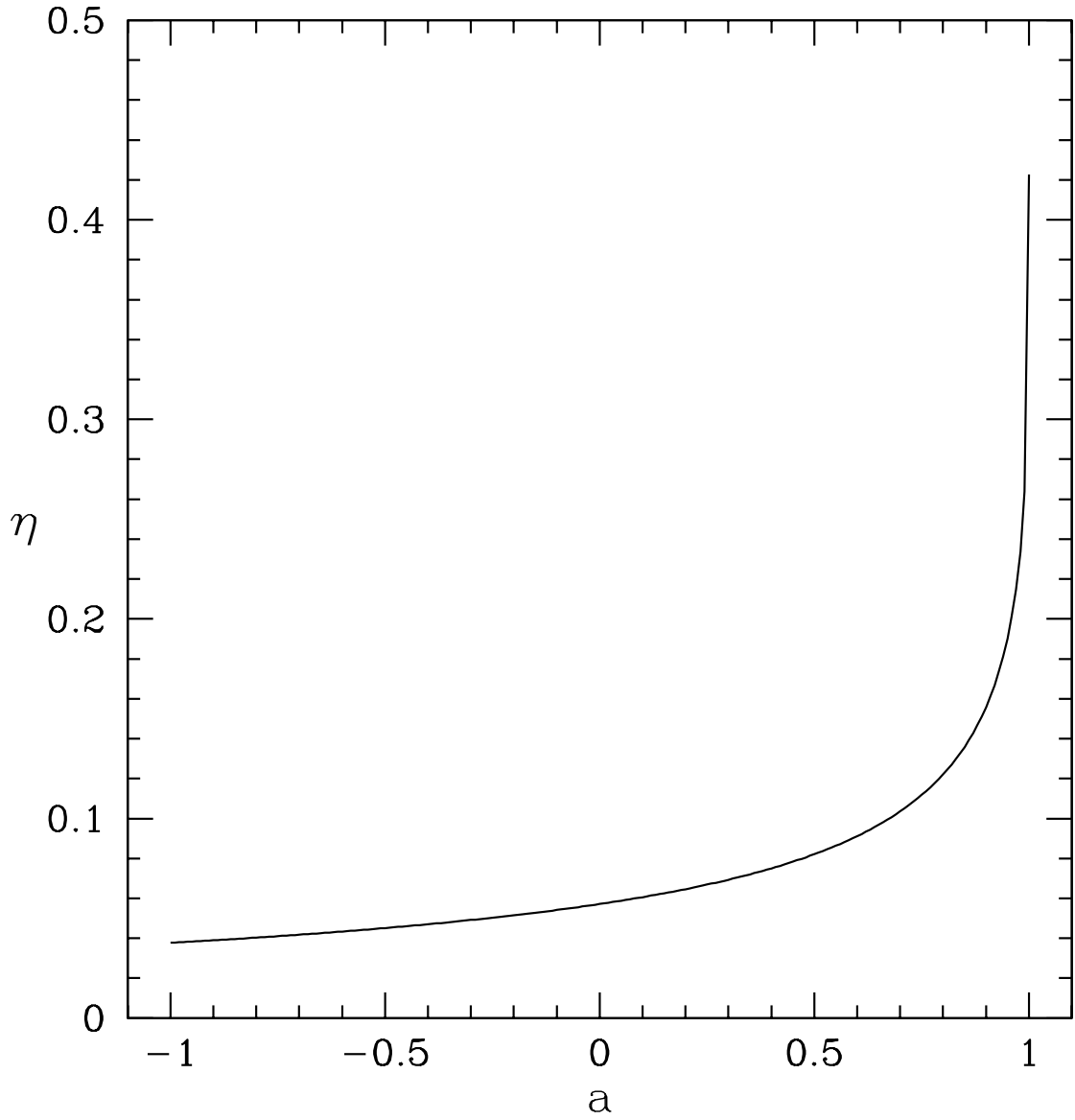}
\caption{Dependence of the efficiency factor $\eta$ on the angular momentum 
of the black hole.}
\end{figure}

\end{document}